\documentclass[reprint,
superscriptaddress,
showpacs,
 amsmath,amssymb,
 aps,
prb,
]{revtex4-1}

\usepackage{graphicx}% Include figure files
\usepackage{dcolumn}% Align table columns on decimal point
\usepackage{bm}% bold math
\usepackage{color}
\def\bs{\boldsymbol}

\begin{document}

%\preprint{APS/123-QED}

\title{Spin-selective electron transfer in quantum dot array}% Force line breaks with \\

\author{Shumpei Masuda*}
\affiliation{QCD Labs, Department of Applied Physics, Aalto University, Aalto 00076, Finland}
\affiliation{College of Liberal Arts and Sciences, Tokyo Medical and Dental University, Ichikawa, 272-0827, Japan}
\author{Kuan Yen Tan}
%\footnote{QCD Labs, Department of Applied Physics, Aalto University, Aalto 00076, Finland}
\affiliation{QCD Labs, Department of Applied Physics, Aalto University, Aalto 00076, Finland}

\author{Mikio Nakahara}
%\affiliation{Department of Physics, Shanghai University, 200444 Shanghai, People's Republic of China}
\affiliation{Department of Mathematics, Shanghai University, 99 Shangda Road, Shangai, 200444, China
}
\affiliation{Research Center for Quantum Computing and Department of Physics, Kindai University, Higashi-Osaka, 577-8502, Japan\\
*masulas@tmd.ac.jp}

%\email{nakahara@math.kindai.ac.jp}

\begin{abstract}
We propose a spin-selective coherent electron transfer in a silicon-quantum-dot array.
Oscillating magnetic fields and temporally controlled gate voltages are utilized to separate the electron wave function into different quantum dots depending on the spin state.
We introduce a non-adiabatic protocol based on $\pi$-pulses and an adiabatic protocol which offer fast electron transfer and 
robustness against the error in the control-field pulse area, respectively.
We also study a shortcut-to-adiabaticity protocol which compromises these two protocols.
We show that this scheme can be extended to multi-electron systems straightforwardly and used for non-local manipulations of electrons.
\end{abstract}

%\pacs{02.30.Yy, 37.90.+, 67.85.Fg, 03.75.Lm}% PACS, the Physics and Astronomy
\maketitle

%\tableofcontents

\section{Introdunction}
Spins in silicon-based quantum dots offer a promising platform for fault-torrelant quantum information processing \cite{Loss1998}.
Fidelities of readout and single-qubit control above the surface code threshold~\cite{Fowler2012} have been demonstrated, courtesy of
exceptionally long lifetimes~\cite{Morello2010,Pla2012,Muhonen2014} and coherence times~\cite{Maune2012,Veldhorst2014,Kawakami2014,Itoh2014}.
These are two figures of merit that are highly desirable for a scalable quantum computing architecture.
Various types of qubit operations have been demonstrated \cite{Noiri2016,Takeda2016,Collard2016} including two-qubit logic gates using the exchange interaction between single spins in isotopically enriched silicon~\cite{Veldhorst2015}.
On the other hand, single electron pumps \cite{Kouwenhoven1991,Blumenthal2007,Jehl2013,Connolly2013,Rossi2014,Pekola2015,Tuomo2015} and the shuttling of single electron \cite{Chan2011,Baart2016} in quantum dots have also been demonstrated at metrological accuracy. 
In fact, single-spin shuttling in a GaAs system quantum dot array has recently been demonstrated using this shuttling operation, and has been shown to preserve the spin coherence up to macroscopic distances \cite{Baart2016}.

In cold atom systems, the coherent transport of neutral atoms \cite{Mandel2003PRL} and the creation of highly-entangled states of neutral atoms  have been demonstrated by utilizing the hyperfine spin-dependent optical lattice potentials \cite{Mandel2003Nat}.
Two-qubit gate operations employing such state-dependent potentials have been studied theoretically \cite{Jaksch1999,Lapasar2011,Lapasar2014}. To the best of our knowledge, however, no spin-selective electron transfer which offers non-local qubit operations in a quantum dot array has been demonstrated.

In this paper, we propose a scheme for spin-selective coherent electron transfer in a quantum dot array achievable using the proven experimental techniques in single-spin shuttling \cite{Chan2011,Baart2016} in a silicon qubit architecture \cite{Veldhorst2015,Takeda2016,Collard2016}.
The gradient of oscillating magnetic fields and controlled gate voltages are utilized to separate the electron wave function into different quantum dots in a spin-selective manner.
This method can be used for quantum non-demolition measurement of electron spin~\cite{Sarovar2008,Puri2014,Muhonen2017} if it is followed by a measurement of the electron position without dissipating the electron.
We propose non-adiabatic and adiabatic protocols.
A simple non-adiabatic transfer based on $\pi$-pulses is fast but also relatively sensitive to the error in timing and amplitude of the control field.
Our adiabatic protocol is based on stimulated Raman adiabatic passage (STIRAP) which is a well-known, efficient protocol for state-to-state population transfer \cite{Gaubatz1990,Bergmann1998,Vitanov2001}.
We introduce the spin-selective STIRAP (spin-STIRAP) which provides robustness against the errors although the operation time is longer than that of the non-adiabatic protocol.
We also examine a non-adiabatic electron transfer based on a shortcut-to-adiabaticity protocol \cite{Torrontegui2013,Masuda_review2016} which is referred to as the invariant-based engineering protocol \cite{Chen2012}. 
It is faster than the spin-STIRAP and more robust against the error of the control field than the $\pi$-pulse protocol.
%We also study the stability of the controls to the potential fluctuation.
Furthermore, we show that this scheme can be extended to multi-electron systems to implement two-qubit gates.
We propose non-local phase manipulations of electrons as an example.

\section{ spin-selective electron transfer}
\begin{figure}
\begin{center}
\includegraphics[width=7.5cm]{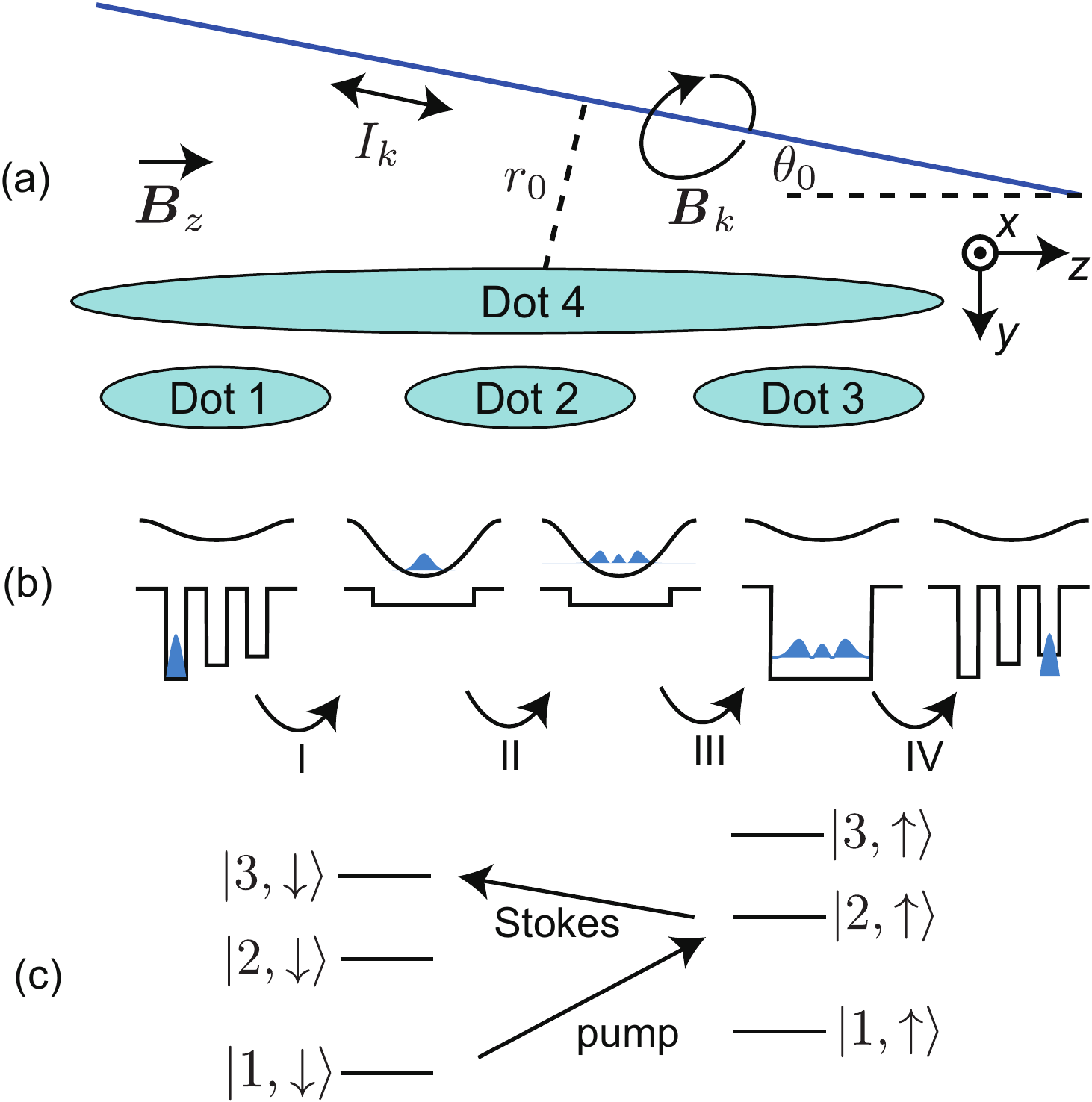}
\end{center}
\caption{(Color online) (a) Schematics of the proposed system. The blue line represents the conducting lead for the AC control currents $I_{k}$ ($k=$ p, S) producing the magnetic fields ${\bm B}_{k}=(B_{k},0,0)$.
The conducting lead is tilted with respect to the dot array by angle $\theta_0$.
Here, $r_0$ is the distance of the lead from the center of Dot 4.
(b) Schematics of the spin-selective transfer of a single electron. The top black curves represent the potential profile of Dot 4, and the bottom black lines represent the potentials of the dot array. The blue color represents the square of the amplitude of the wave function of the spin-down electron.
(c) Energy diagram of the system in step II.
Here, $|i,\uparrow(\downarrow)\rangle$ is the $i$th instantaneous eigenstate trapped in Dot 4 with spin up (down).
The AC magnetic field ${\bm B}_{\rm p}$ (pump field) couples $|1,\downarrow\rangle$ and $|2,\uparrow\rangle$,
and ${\bm B}_{\rm S}$ (Stokes field) couples $|2,\uparrow\rangle$ and $|3,\downarrow\rangle$.}
\label{dot_leads_7_17_16}
\end{figure}

%We study spin-selective electron transfer in an array of two-dimensional quantum dots.
We first consider a four-dot system shown in Fig.~\ref{dot_leads_7_17_16}(a), where
three quantum dots align along the $z$-axis, and a wider quantum dot is located parallel to the array in the $yz$-plane.
The heights of the potential barriers between the dots and the depths of the potential wells are tunable.
There is a stationary uniform magnetic field ${\bm B}_{z}=(0,0,B_{z})$ parallel to the two-dimensional electron gas.
A conducting lead carries the AC currents, $I_{k}$ ($k=$ p, S), which induce the AC magnetic fields, ${\bm B}_{k}=(B_{k},0,0)$, perpendicular to the two-dimensional electron gas.
The conducting lead is separated from the center of Dot~4 by distance $r_0$, and is tilted with respect to the dot array by angle $\theta_0$ to enhance the influence of the spatial dependence of the magnetic field ${\bm B}_{k}$ on the electron in the quantum dots.
This spatial dependence of the magnetic field plays an essential role in our scheme.

We consider spin-selective electron transfer from Dot~1 to Dot~3, in which only a spin-down electron is transferred, while a spin-up electron returns to Dot 1 in the end of the control.
We assume that the electron is initially trapped in Dot 1 and that the state of the electron is a superposition of the lowest-energy spin-up state and the lowest-energy spin-down state.

The electron transfer protocol is illustrated in Fig.~\ref{dot_leads_7_17_16}(b). It consists of four steps:
(I)~adiabatic transfer of the electron to Dot 4,
(II)~non-adiabatic or adiabatic spin-selective level transfer, 
(III)~adiabatic transfer of the electron to the in-line dots (Dots 1, 2, 3),
(IV)~loading of the spin-down electron to Dot 3.

In step I, the gate voltages of the in-line dots are gradually increased, and the potential well of Dot 4 is deepened so that the electron is transferred to Dot 4 without energy excitations.
In step II, the lowest energy state with spin down is transferred to the second spin-down excited state,
while the spin-up electron remains in the spin-up ground state.
We use Dot 4 for the level transfer instead of Dot 1 nor the combined dot composed of the in-line dots in order to enhance the influence of the AC magnetic fields to the electron and also to reduce the influence of the fluctuation of the gate voltage at the in-line dots.
Three different methods for step II are introduced later.
In step III, the barriers of the in-line dots are lowered and the potential depth of Dot 4 is reduced so that the electron is adiabatically transferred to the combined dot.
In step IV,  the barriers of the in-line dots are gradually increased and the depths of the potential wells are tuned so that the Dot 3 has the highest potential among the in-line dots.
In the end of step IV, the spin-down electron is adiabatically carried into Dot 3 because the wave function of the second excited instantaneous eigenstate with spin down is located in Dot~3.
On the other hand, the spin-up electron returns to Dot 1.
Note that the state of the electron is the superposition of these two states. 
This method followed by a measurement to determine the dot where the electron is trapped \cite{Elzerman2004} can be used for quantum non-demolition measurement of the electron spin because a position measurement projects the electron state to spin-up or spin-down. 

The adiabatic transfer of electrons between quantum dots have been routinely used.
Therefore we mainly discuss step II in the following.
The duration of step II is much longer than the other steps. 
Thus, step II dominates the execution time of the transfer process.
Details of step~II are analyzed in Sec.~\ref{Non-adiabatic and adiabatic spin-selective level transfers} and
 Sec.~\ref{Step II}.
Details of step~IV are shown in Sec.~\ref{Step IV}, while details of steps~I and III are given in Appendix~\ref{Step I and step III}. 

Simpler spin-selective electron transfers using only the in-line dots without Dot 4 might be possible if the fluctuation of the gate voltages of the dots and barriers are negligible.
However the fluctuation causes unwanted fluctuation of the separation of the energy levels and thus lowers the transfer efficiency.
On the other hand, in our scheme, the electron is prepared in a single dot for adiabatic loading to a selected dot.
Thus, we can restrain the influence of the potential fluctuation.

\section{Non-adiabatic and adiabatic spin-selective level transfers}
\label{Non-adiabatic and adiabatic spin-selective level transfers}
In step~II, the electron is trapped in Dot~4.
To detail the schemes of step~II, we use the energy eigenstates in Dot~4 for ${\bm B}_{k}=0$ $(k=\mbox{p,\ S})$ as a basis of the system.
The energy diagram of the system is illustrated in Fig.~\ref{dot_leads_7_17_16}(c).
Here, $|i,\downarrow(\uparrow)\rangle$ for $i=1,2,3$ denote the first three lowest energy levels with spin down (up)
in the $z$-direction.
The energy separations of the levels are nonuniform because the potential of Dot 4 is anharmonic.
The stationary magnetic field ${\bm B}_{z}$ causes the Zeeman splitting with energy difference $g\mu_{\rm B}B_z/\hbar$ between spin-up and -down states, where $g$ is the electron $g$-factor and $\mu_{\rm B}$ is the Bohr magneton. 

We aim at a spin-selective transfer in which only spin-down electron is transferred from $|1,\downarrow\rangle$ to $|3,\downarrow\rangle$ , while spin-up electron is unaffected.
The frequencies $\omega_{k}$ of ${\bm B}_{k}$ are tuned to be $\omega_{\rm p}=(E_{2,\uparrow}-E_{1,\downarrow})/\hbar$ and $\omega_{\rm S}=(E_{3,\downarrow}-E_{2,\uparrow})/\hbar$, where $E_{i,\uparrow(\downarrow)}$ is the energy eigenvalue of $\\*|i,\uparrow(\downarrow)\rangle$, so that the pump field ${\bm B}_{\rm p}$ couples $|1,\downarrow\rangle$ and $|2,\uparrow\rangle$, and the Stokes field ${\bm B}_{\rm S}$ couples $|2,\uparrow\rangle$ and $|3,\downarrow\rangle$.
Note that $|1,\uparrow\rangle$ is not coupled to the other states by ${\bm B}_{k}$.

We represent the effective Hamiltonian of the system using, as a basis, the subset of states \\$\{ |1,\downarrow\rangle, |2,\uparrow\rangle,  |3,\downarrow\rangle \}$ coupled by the resonant magnetic fields.
%We assume that the other excited states do not contribute to the dynamics of the system at low temperature because the oscillating fields do not couple them to the subset of states above.
The pulsed magnetic fields used in step II are represented as
\begin{eqnarray}
B_{k}(t,{\bm r})=B_{k}^{\rm (e)}(t)\eta({\bm r})\cos(\omega_{k} t)
\label{Bx1}
\end{eqnarray}
with the envelope function $B_{k}^{\rm (e)}(t)$, which is the envelope of the pump and the Stokes fields at the center of Dot 4.
Here, $\eta({\bm r})$ is the ratio of the intensity of the field at ${\bm r}$ to $B_{k}^{\rm (e)}$.  
Thus, $\eta({\bm r})$ characterizes the spatial dependence of the magnetic field.
Using the rotating frame and the rotating wave approximation (RWA), the Hamiltonian of the three-level system can be put in the form
\begin{eqnarray}
H_{\rm RWA}(t) = \frac{\hbar}{2}\left( \begin{array}{ccc}
0 & \Omega_{\rm p}(t) & 0  \\
\Omega_{\rm p}(t) & 0 & \Omega_{\rm S}(t)  \\
0 & \Omega_{\rm S}(t) & 0
\end{array}
 \right),
 \label{HRWA1}
 \end{eqnarray}
with the Rabi frequencies given by
\begin{eqnarray}
\Omega_{k}(t) &=& \frac{B_{k}^{\rm (e)}(t) g \mu_{\rm B} \mu_{k}}{2\hbar},
\label{eqRabi_5_18_17}
\end{eqnarray}
and the overlapping factors defined by
\begin{eqnarray}
\mu_{\rm p} &=& \int d{\bm r} \phi_{1}^\ast({\bm r}) \eta({\bm r}) \phi_{2}({\bm r}),\nonumber\\
\mu_{\rm S} &=& \int d{\bm r} \phi_{2}^\ast({\bm r}) \eta({\bm r}) \phi_{3}({\bm r}),
\label{mu_10_26_17}
\end{eqnarray}
where $ \phi_{i}({\bm r})=\langle{\bm r},\downarrow(\uparrow)|i,\downarrow(\uparrow)\rangle$
(see Appendix~\ref{HRWA_ap} for details of the derivation of $H_{\rm RWA}$).
If the magnetic fields ${\bm B}_{k}$ were spatially uniform, they could not couple the energy levels because the energy eigenvectors are orthogonal to each other.
The spatial dependence of the magnetic fields realizes the coupling between the energy levels.

It has been demonstrated that the valley separation can be tuned via electrostatic gate control of quantum dots providing the splittings spanning 0.3--0.8~meV \cite{Yang2013}.
Thus, we assume that other valley states are located sufficiently above $|i,\uparrow(\downarrow)\rangle$, and multi-valley relaxation effects are negligible. We also assume that the relaxation rate from the excited states $|2,\uparrow\rangle, |3,\downarrow\rangle$ to the lower energy states induced by the interaction with other electrons is small enough compared to the duration of step II.

\subsection{Non-adiabatic spin-selective electron transfer based on simple $\pi$-pulse control}
One of the non-adiabatic schemes of step II is composed of a $\pi$-pulsed pump field followed by a $\pi$-pulsed Stokes field that are separated from each other in the time domain as depicted in Fig.~\ref{rabi_com}(a).
The envelope functions are given as
\begin{eqnarray}
B_{k}^{\rm (e)}(t) = 
\begin{cases}
B_{k}^{0} &  {\rm for} \  |t-T_{k}| \le \tau_{k}/2,\\
 0 &  {\rm for} \  |t-T_{k}| > \tau_{k}/2,
\end{cases}
\end{eqnarray}
where $B_{k}^{0}$ is the amplitude of the rectangular pulse and $\tau_{k}$ is the pulse width.
%The pump pulse drives the state from $|1,\downarrow\rangle$ to $|2,\uparrow\rangle$, and then the Stokes pulse 
%drives the state from $|2,\uparrow\rangle$ to $|3,\downarrow\rangle$.
When the pulse areas are $\pi$, that is, 
\begin{eqnarray}
%g\mu_{\rm B}B_{p/S}^{(0)} \mu_{p/S}\tau_{p/S} = \pi,
B_{k}^{0} = \frac{2\pi\hbar}{g\mu_B \mu_{k}\tau_{k}},
\end{eqnarray}
the state is transferred from $|1,\downarrow\rangle$ to $|2,\uparrow\rangle$ due to the pump pulse, and then transferred from $|2,\uparrow\rangle$ to $|3,\downarrow\rangle$ due to the Stokes pulse.
However this scheme is sensitive to the error in the pulse area compared to the adiabatic scheme discussed below. 
Inaccuracy of the pulse area causes the imperfection of the population transfer.
Another method of wave function splitting depending on electron spin based on $\pi$-pulse control is discussed in Appendix~\ref{Another method of wave function splitting}.

%The Hamiltonian of the system is written as
%\begin{eqnarray}
%H = \frac{{\bm p}^2}{2m} + V({\bm x}) +  \frac{g\mu_{\rm B}}{\hbar}{\bm B}\cdot{\bm S}
%\end{eqnarray}
%with $g$ electron g-factor and $\mu_{\rm B}$ the Bohr magneton.
%The magnetic field is given by
%\begin{eqnarray}
%{\bm B} = {\bm B}_{\rm z} + {\bm B}_{\rm x}(t)
%\end{eqnarray}
%with $B_{\rm z} = (0,0,B_z)$ and $B_{\rm x} = (B_x(t),0,0)$.
%Because of the different gate voltages the energy intervals of the eigenstates are different.

\subsection{Spin-selective STIRAP}
Stimulated Raman adiabatic passage (STIRAP) has been widely studied for population transfer of molecules \cite{Coulston1994, Martin1996, Halfmann1996, Malinovsky1997, Kobrak1998b, Kurkal2001a, Cheng2007, Jakubetz2012}, 
transport of single atoms \cite{Eckert2004,Eckert2006,Opatrny2009,OSullivan2010,Morgan2011,Morgan2013}, electrons \cite{Greentree2004,Jong2009} and BECs \cite{Graefe2006,Rab2008,Nesterenko2009,Rab2012}.
The remarkable properties of this protocol have already been demonstrated in diverse
areas such as chemical reaction dynamics \cite{Dittmann1992}, laser-induced cooling of atomic gases \cite{Kulin1997}, light beams propagating in three evanescently coupled optical waveguides \cite{Longhi2007,Lahini2008,Menchon2012,Menchon2013}, sound propagation in sonic crystals \cite{Menchon2014}
and control of a superconducting qubit \cite{Kumar2016}.
In spin-based quantum computing architecture, this protocol can be utilized to transfer qubits coherently across large distances \cite{Hollenberg2006}. 

\begin{figure}
\begin{center}
\includegraphics[width=7cm]{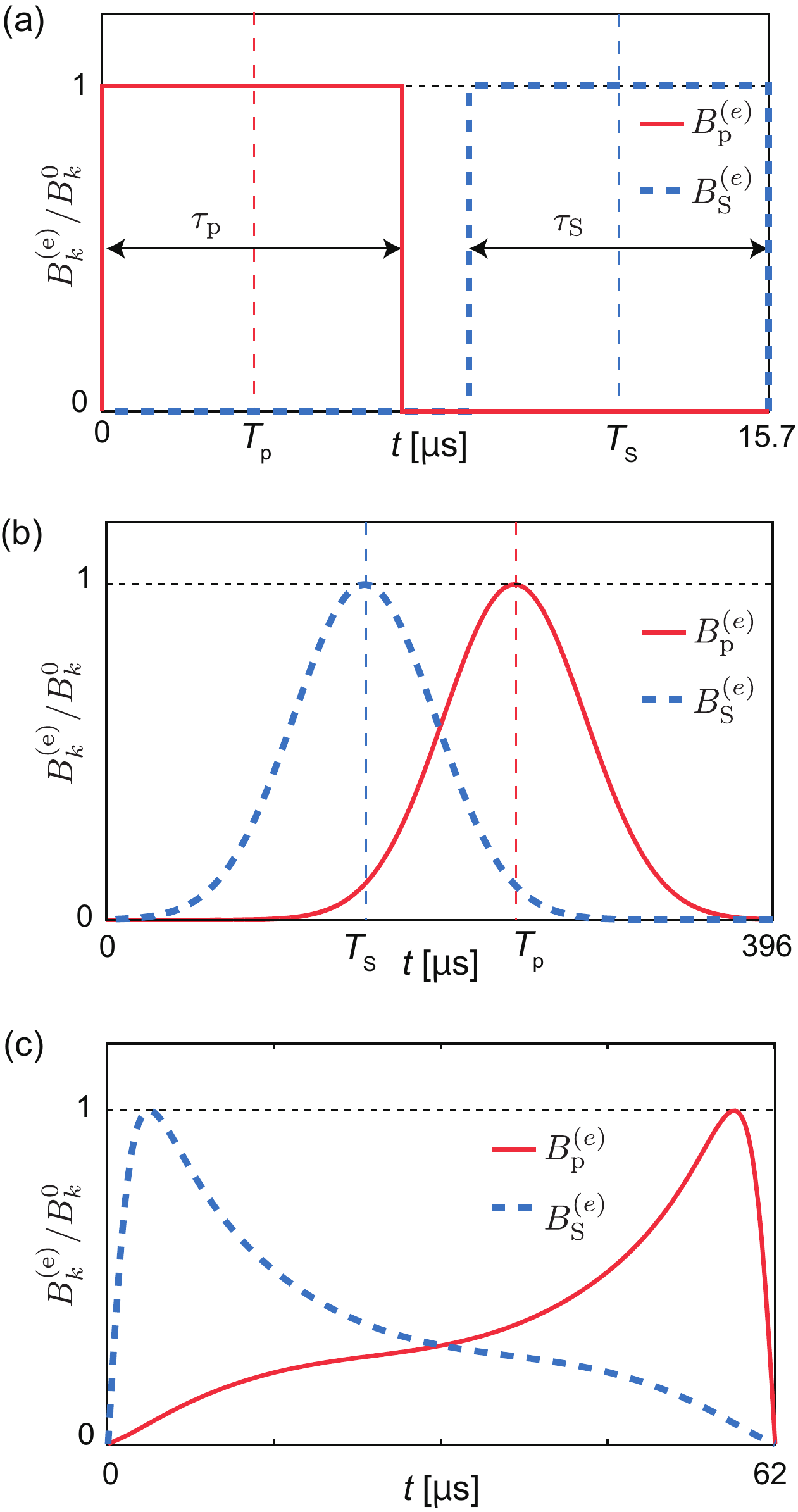}
\end{center}
\caption{(Color online) Envelop function of the magnetic fields $B_{\rm p}^{\rm (e)}$ and $B_{\rm S}^{\rm (e)}$ normalized by $B_{\rm p}^{0}$ and $B_{\rm S}^{0}$, respectively,  for $T_1(=0)<t<T_2$ for
(a) $\pi$-pulse control, (b) spin-STIRAP and (c) invariant-based engineering protocol.
$B_{k}^{0}$ is the maximum intensity of the pules. The parameters used are shown in Table~\ref{table_STEP2}.
}
\label{rabi_com}
\end{figure}

Here, we introduce the spin-STIRAP of an electron in Dot 4. The envelope functions of the pump and the Stokes fields are represented as
\begin{eqnarray}
B_{k}^{\rm (e)}(t) = 
B_{k}^{0} \exp\Big{[}-\frac{(t-T_{k})^2}{2\sigma^2}\Big{]},
\end{eqnarray}
where $\sigma$ is defined by
\begin{eqnarray}
\sigma = \frac{\rm FWHM}{2\sqrt{2\ln2}}
\end{eqnarray}
with the full width at half-maximum (FWHM) and the maximum intensity $B_{k}^{0}$ of the peak centered at $T_{k}$.
The separation of the peaks is chosen as
\begin{eqnarray}
T_{\rm p} - T_{\rm S} = \frac{3 {\rm FWHM}}{4\sqrt{\ln2}}.
\end{eqnarray}
Note that the pump pulse follows the Stokes pulse, $T_{\rm p} > T_{\rm S}$, as shown in Fig.~\ref{rabi_com}(b).

A time-dependent, field-dressed eigenstate of the system, which is a linear combination of the field-free states, is represented as
\begin{eqnarray}
|\phi_0(t)\rangle = \cos\Theta(t) |1,\downarrow\rangle - \sin\Theta(t) |3,\downarrow\rangle,
\end{eqnarray}
where $\Theta(t)$ is given by
\begin{eqnarray}
\tan\Theta(t) = \frac{\Omega_{\rm p}(t)}{\Omega_{\rm S}(t)}.
\end{eqnarray}
Because the Stokes pulse precedes the pump pulse, $\Omega_{\rm p}\ll \Omega_{\rm S}$ and $|\phi_0\rangle= |1,\downarrow\rangle$ at the initial time, $T_1$, of step~II; 
$\Omega_{\rm p}\gg \Omega_{\rm S}$ and $|\phi_0\rangle= |3,\downarrow\rangle$ at the final time, $T_2$, of step~II.
The STIRAP control is robust against the change in the profile of the  $B_{k}^{\rm (e)}$.

\subsection{Shortcuts to adiabaticity}
\label{Shortcuts to adiabaticity}
Assisted adiabatic transformation or shortcut-to-adiabaticity (STA) protocols have been developed to generate the same target state as reference adiabatic dynamics, with overall weaker driving fields and/or in a shorter time \cite{Torrontegui2013,Masuda_review2016}. 
The STA protocols have been utilized for manipulations of, e.g., isolated atoms and molecules \cite{cd1,cd3,Masuda2014,Du2016,An2016}, spin systems \cite{Berry2009,Campo2012,Fasihi2012,Takahashi2013}, Bose-Einstein condensates \cite{Masuda2009,Muga2009,Schaff2011,Bason2011,Masuda2012,Torrontegui2012,Deffner2014,Masuda2014b} and electron spin of a single nitrogen-vacancy center in diamond \cite{Zhang2013,Zhou2016}.
Several STA protocols have been applied to STIRAP systems, for example Loop STIRAP \cite{Uanyan1997}, counter-diabatic \cite{cd1,cd3}, fast-forward \cite{Masuda2015a} and invariant-based engineering protocols \cite{Chen2012,Chen2012b}.

We show that the STA protocol can be used for the spin-selective transfer faster than the STIRAP control and more robust than the $\pi$-pulse control using the invariant-based engineering protocol \cite{Chen2012b}.
Using the result in Sec. III of Ref.[\!\!\citenum{Chen2012b}] and Eq.~(\ref{eqRabi_5_18_17}) of this paper we can derive the magnetic fields as
\begin{eqnarray}
B_{\rm p}^{\rm (e)} (t) = \frac{4 \hbar(\dot{\beta}\cot{\gamma}\sin{\beta} + \dot{\gamma}\cos{\beta})}{g\mu_{\rm B}\mu_{\rm p}},\nonumber\\
B_{\rm S}^{\rm (e)} (t) = \frac{4 \hbar(\dot{\beta}\cot{\gamma}\cos{\beta} - \dot{\gamma}\sin{\beta})}{g\mu_{\rm B}\mu_{\rm S}},
\label{B_LR_10_18_17}
\end{eqnarray}
where
\begin{eqnarray}
\gamma = \sum_{j=0}^4 a_j t^j,\nonumber\\
\beta = \sum_{j=0}^3 b_j t^j,
\label{gamma_10_18_17}
\end{eqnarray}
with $a_0=\varepsilon$, $a_1=0$, $a_2=16(\delta-\varepsilon)/T_f^2$, $a_3=-32(\delta-\varepsilon)/T_f^3$,
$a_4=16(\delta-\varepsilon)/T_f^4$,
$b_0=b_1=0$, $b_2=3\pi/(2T_f^2)$ and $b_3=-\pi/(T_f^3)$.
Here, $T_f=T_2 - T_1$ is the duration of the control; 
We have $\gamma(0)=\gamma(T_f)=\varepsilon$, $\beta(0)=0$ and $\beta(T_f)=\pi/2$.
Figure~\ref{rabi_com}(c) shows the time dependence of $B_{\rm p}^{\rm (e)}$ and $B_{\rm S}^{\rm (e)}$ for the parameters in Table~\ref{table_STEP2}(c).

A relevant dynamics of the system governed by $H_{\rm RWA}$ in Eq.~(\ref{HRWA1}) with $B_{\rm p}^{\rm (e)}$ and $B_{\rm S}^{\rm (e)}$ in Eq.~(\ref{B_LR_10_18_17}) is explicitly writen as 
\begin{eqnarray}
|\varphi_0(t)\rangle = \left( \begin{array}{c}
\cos\gamma(t)\cos\beta(t) \\
-i\sin\gamma(t) \\
-\cos\gamma(t)\sin\beta(t)
\end{array}
 \right),
 \label{phiLR_10_18_17}
 \end{eqnarray}
 which is an instantaneous eigenstate of the invariant 
 \begin{eqnarray}
 I(t)=\frac{\hbar}{2}\Omega_0\left( \begin{array}{ccc}
0 & \cos\gamma\sin\beta & -i\sin\gamma \\
\cos\gamma\sin\beta & 0 & \cos\gamma\cos\beta \\
i\sin\gamma & \cos\gamma\cos\beta & 0
\end{array}
 \right),\nonumber\\
  \end{eqnarray}
which satisfies $dI/dt=0$.
$\Omega_0$ is an arbitrary constant with unit of frequency.
Importantly, if $\varepsilon$ is sufficiently small, we have $|\varphi_0(0)\rangle \simeq |1,\downarrow\rangle$ and $|\varphi_0(T_f)\rangle \simeq - |3,\downarrow\rangle$.
Therefore the magnetic field in Eq.~(\ref{B_LR_10_18_17}) can approximately drive the initial state $|1,\downarrow\rangle$ to target state $|3,\downarrow\rangle$ up to overall phase.
Because $|1,\downarrow\rangle$ and $|3,\downarrow\rangle$ are not exactly $|\varphi_0(0)\rangle$ and $-|\varphi_0(T_f)\rangle$, the fidelity of this control is slightly less than unity.

As shown in Sec.~\ref{Numerical results}, $T_f$, $\varepsilon$ and $\delta$ determine the profile of $B_{k}^{\rm (e)}$;
$\varepsilon$ also determines the fidelity when there is no noise; $\delta$ determines the population of intermediate state during the control and the fidelity of the control.

\section{Numerical results}
\label{Numerical results}
The duration of step~II is much longer than those of other steps. 
Thus it dominates the total duration of the process.
We first study the efficiency of step~II then show the numerical results for step IV.
The numerical results for steps~I and~III are shown in Appendix~\ref{Step I and step III}.

\subsection{Step II}
\label{Step II}
\begin{figure}
\begin{center}
\includegraphics[width=6cm]{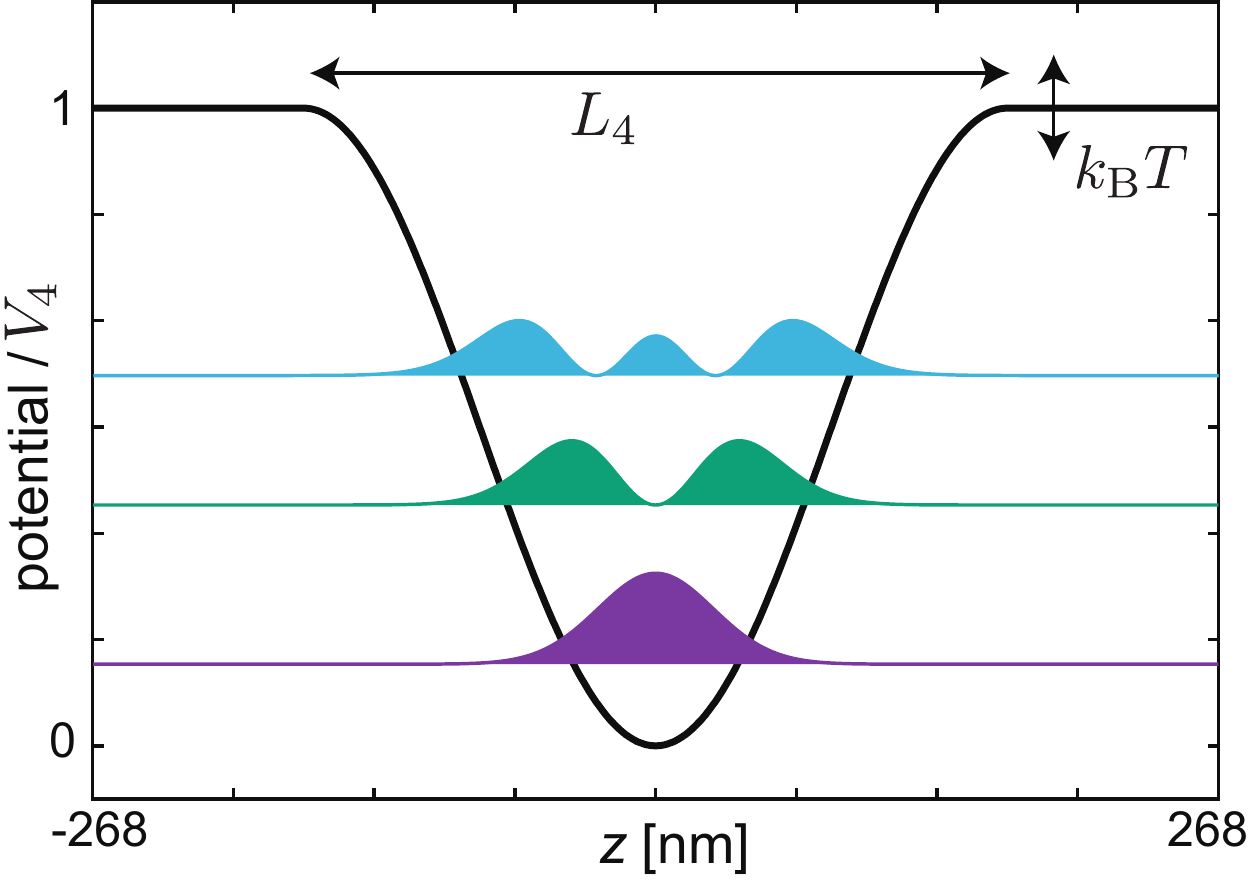}
\end{center}
\caption{(Color online) Potential of Dot~4 normalized by $V_4$. 
The colors show the square of the amplitude of the wave functions of the three lowest levels with either spin up or down.
The vertical arrows depict the fluctuation of $V_4$ on the order of $k_{\rm B}T$.}
\label{potential_sin}
\end{figure}

We examine the efficiency of the level transfer in step~II using a one-dimensional model for Dot~4 illustrated in Fig.~\ref{potential_sin} with the Hamiltonian:
\begin{eqnarray}
H = \frac{{p}^2}{2m^*} + V({z}) +  \frac{g\mu_{\rm B}}{\hbar}{\bm B}\cdot{\bm S},
\label{Hami_10_9_16}
\end{eqnarray}
where $m^*$ is the effective electron mass and ${\bm S}$ is the electron spin. We assume that the confinement of the electron in the $y$-direction is at least twice stronger than the confinement in the $z$ direction.
Then, the 1-dimensional model can approximate a few of the lowest energy eigenstates of the dot that we utilize.
We take the potential of Dot 4 as
\begin{eqnarray}
V(z) = 
\begin{cases}
V_4  \sin^2{(}{\pi  z}/{ L_4}{)} &  {\rm for} \ |z| \le L_4/2,\\
V_4 &  {\rm for} \ |z| > L_4/2.
\end{cases}
\label{eqV4}
\end{eqnarray}
Here, $L_4$ is the width of the dot, and the depth of the potential well is $V_4$.
%The potential barriers $V_{{\rm b}i}$ are adiabatically decreased in STEP1 for $0\le t < T_1$ from $6V_{\rm b}^{(0)}$ 
%to $V_{\rm b}^{(0)}$ as
%\begin{eqnarray}
%V_{{\rm b}i} = V_{\rm b}^{(0)} [6-5R(t)]
%\end{eqnarray}
%with $R(t) = [t-\sin(\omega t)/\omega]/T_1$ and $\omega = 2\pi/T_1$.
%All the potential barriers are taken to be the same for the simplicity.
The square of the amplitude of the wave functions of the three lowest levels with either spin up or down are shown in Fig.~\ref{potential_sin}.
%\begin{figure}
%\begin{center}
%\includegraphics[width=6cm]{potential_ph2_wf_v2.pdf}
%\end{center}
%\caption{(Color online) Schematics of the one-dimensional model system. The inset shows the amplitude of the wave function of the instantaneous eigenstates.
%The height of potential barrier and the depth of the potential wells are denoted by $V_{\rm bi}$ and $V_{\rm dj}$ for $i=1,2,3,4$ and $j=1,2,3$, respectively.
%The width of the dots $L_{d}$ and the width of the potential barriers $L_{b}$ are the same.}
%\label{potential_ph_4}
%\end{figure}

\begin{figure}
\begin{center}
\includegraphics[width=7cm]{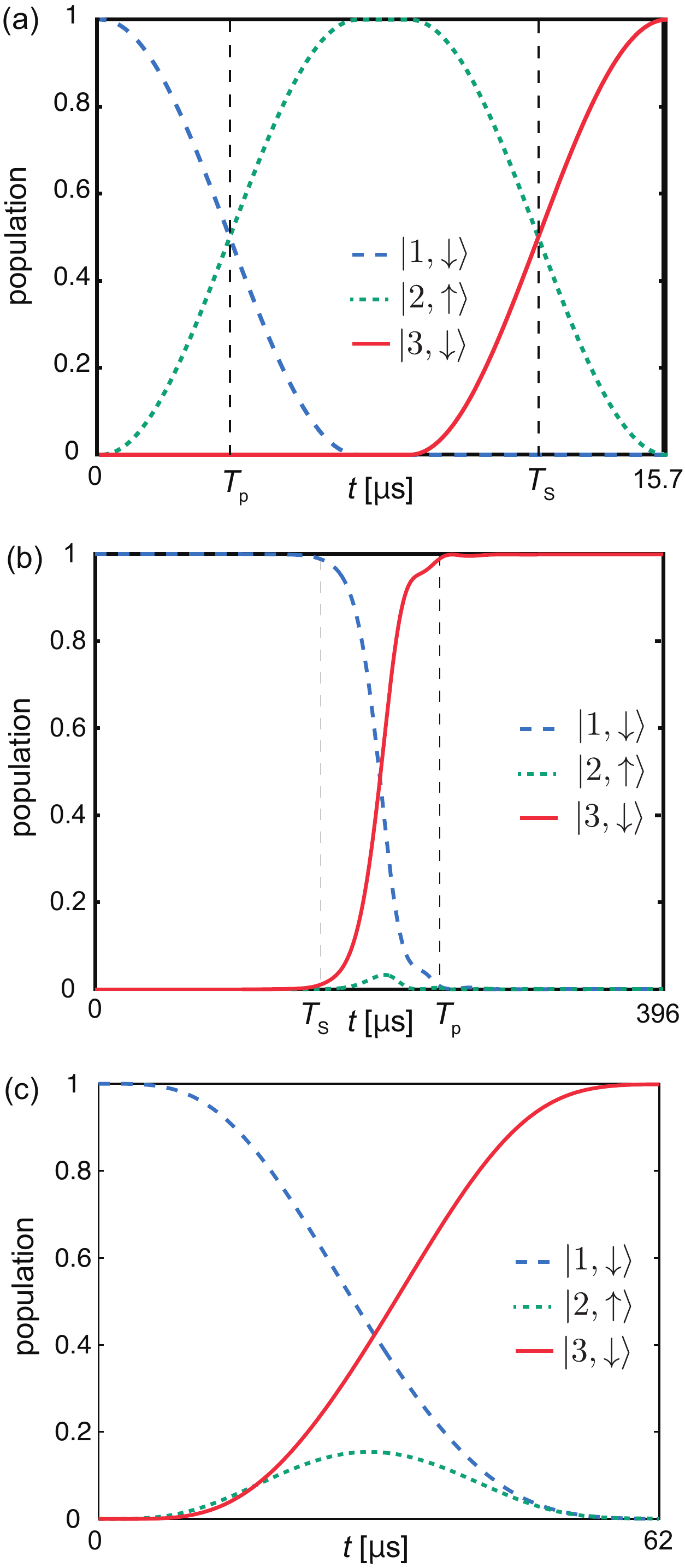}
\end{center}
\caption{(Color online) Time-dependence of the populations in (a) $\pi$-pulse control, (b) spin-STIRAP, (c) invariant-based engineering protocol. The parameters used are shown in Table~\ref{table_STEP2}.}
\label{pop}
\end{figure}

%The initial state is the energy eigenstate $|1,\downarrow\rangle$.
The dynamics of the system in step~II is simulated using the three-level model
expanded by $|1,\downarrow\rangle, |2,\uparrow\rangle, |3,\downarrow\rangle$ for $T_1\le t\le T_2$ without RWA.
In the numerical simulation, we solve the time-dependent Schr\"odinger equation with a fourth-order Runge-Kutta integrator with the time step of approximately 1~ps.
%, where $T_1$ and $T_2$ are the initial and final time of STEP~II. 
The frequencies of the AC magnetic fields are $f_{\rm p}=\omega_{\rm p}/(2\pi)\simeq$ 47 GHz and $f_{\rm S}=\omega_{\rm S}/(2\pi)\simeq$ 32 GHz
for $g=2$, $L_4 =335$~nm, $V_4=0.72$~meV, $B_z=0.2$~T and $m^* =0.28~m_e$, where $m_e$ is the electron mass. 
The overlapping factors are $\mu_{\rm p}=-0.05$ and $\mu_{\rm S}=-0.078$ corresponding to $r_0=1.4~L_{\rm 4}$ and $\theta_0\simeq \pi/3 $.
The eigenenergies are calculated by using the Hamiltonian~(\ref{Hami_10_9_16}).
%The middle panels in Figs.~\ref{pop}(a) and \ref{pop}(b) show the time-evolution of the amplitude of the wave function for the non-adiabatic transfer and the STIRAP transfer for the parameter set in Tables~\ref{table_STEP2}(a) and \ref{table_STEP2}(b), respectively.
Figure~\ref{pop}(a) shows the time-dependence of the populations in $|1,\downarrow\rangle$, $|2,\uparrow\rangle$ and $|3,\downarrow\rangle$ in the non-adiabatic spin-selective electron transfer with $\pi$-pulse fields, where
the state is driven to $|2,\uparrow\rangle$ and subsequently to $|3,\downarrow\rangle$.
The parameters used are shown in Table~\ref{table_STEP2}(a).
Figure~\ref{pop}(b) shows the populations under the spin-STIRAP for the parameters in Table~\ref{table_STEP2}(b).
The population is almost directly transferred to $|3,\downarrow\rangle$.
The finite population of $|2,\uparrow\rangle$ around $t=(T_{\rm p}+T_{\rm S})/2$ is due to the finite pulse area.
The duration of the control, $T_2-T_1$, for non-adiabatic transfer is 25 times shorter than that of the spin-STIRAP.
Figure~\ref{pop}(c) shows the populations in the invariant-based engineering protocol.
The population of $|2,\uparrow\rangle$ during the control depends on $\delta$.
The duration of the control, $T_2-T_1$, is 6 times shorter than that of the spin-STIRAP.
The same value of $B_{k}^{0}$ was used for these protocols.
\begin{table}
(a) $\pi$ pulse control\\
\begin{tabular}{|c|c|c|ccccc}
\hline
$T_2-T_1$  & $T_{\rm S}-T_{\rm p}$ &  $\tau_{\rm p, S}$     \\
\hline
15.68 $\mu$s & 8.55 $\mu$s &  7.13 $\mu$s     \\
\hline    
\end{tabular}\\
(b) Spin-STIRAP\\
\begin{tabular}{|c|c|c|ccccc}
\hline
$T_2-T_1$ & $T_{\rm p}-T_{\rm S}$  &   FWHM   \\
\hline
396 $\mu$s & 80 $\mu$s &  88 $\mu$s    \\
\hline   
\end{tabular}\\
(c) Invariant-based engineering protocol\\
\begin{tabular}{|c|c|c|ccccc}
\hline
$T_2-T_1$   & $\delta$ & $\varepsilon$\\
\hline
62 $\mu$s  & $\pi/8$ & 0.02\\
\hline   
\end{tabular}
\caption{Parameters of the pulse fields for the spin-selective, inter-level population transfer in step II.
Other parameters are given as  $B_{\rm p}^0=$ 0.1 mT, $B_{\rm S}^0=$ 0.064 mT, $f_{\rm p}=$ 47 GHz  and $f_{\rm S}=$ 32 GHz. 
The interval between the pump and Stokes pulses in the $\pi$-pulse control was set to be 1.42~$\mu$s to avoid unwanted overlap of the pulses.}
\label{table_STEP2}
\end{table}

%The large overlap factors allows shorter time STIRAP with high fidelity.
%In Fig.~\ref{ovs}(a) and (b), $g_{12}$ and $g_{23}$ are shown as functions of $\theta$ and $r_0$.
%\begin{figure}
%\begin{center}
%\includegraphics[width=6cm]{ovs.jpg}
%\end{center}
%\caption{(Color online) $g_{12}$ and $g_{23}$ for the parameter set in Table.}
%\label{ovs}
%\end{figure}

\begin{figure}[h!]
\begin{center}
\includegraphics[width=5.5cm]{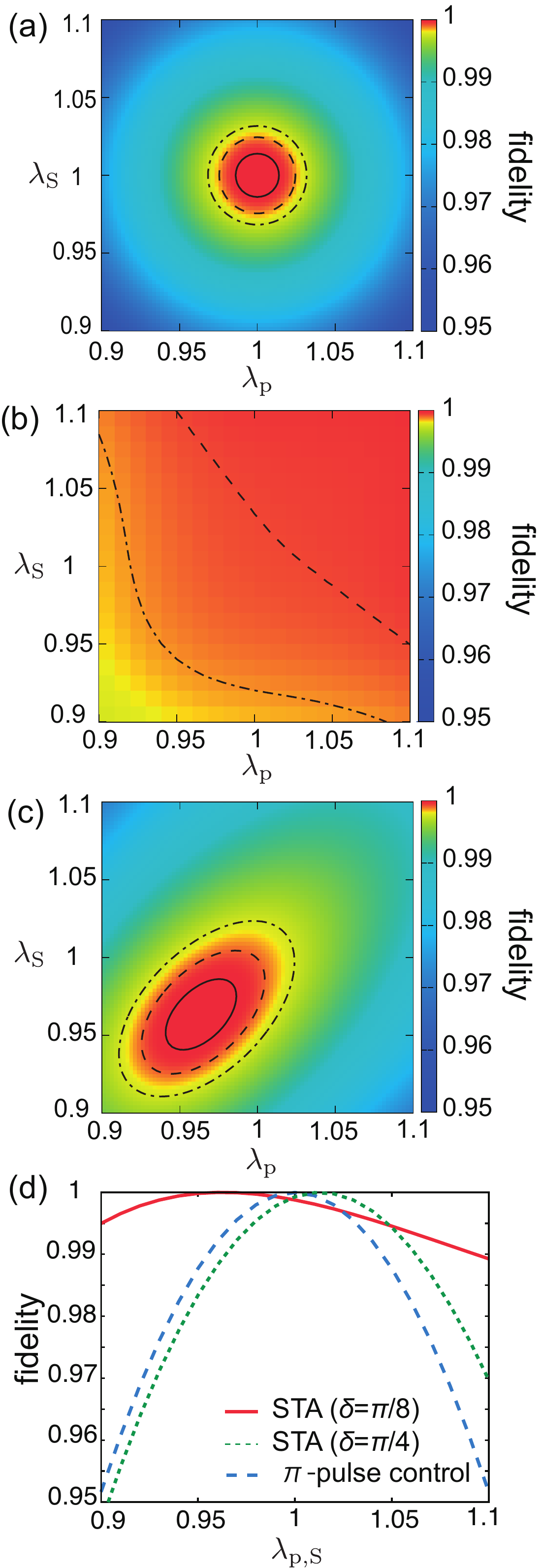}
\end{center}
\caption{(Color online) Dependence of the fidelity defined by the population of $|3,\downarrow\rangle$ at $t=T_2$ on $\lambda_{k}$ for (a) $\pi$-pulse control, (b) spin-STIRAP, (c) invariant-based engineering protocol.
In panels (a)--(c), contours are shown for the values of fidelity, 0.9995 (solid curve), 0.999 (dashed curve), 0.9985 (dashed--dotted curve), while the contour for fidelity 0.9995 is out of the range of panel (b).
(d) The $\lambda_{\rm p}$ dependence of the fidelity of the $\pi$-pulse control and the invariant-based engineering protocol (STA), where $\lambda_{\rm p} =\lambda_{\rm S}$.
The parameters used are shown in Table~\ref{table_STEP2}.
The green dotted curve and the red solid curve correspond to the STA with $\delta=\pi/4$ and $\delta=\pi/8$, respectively.
The blue dashed curve corresponds to the $\pi$-pulse control. }
\label{stability}
\end{figure}

Non-adiabatic transfer schemes take shorter time than the STIRAP transfer scheme.
However the STIRAP transfer is much more robust against the error of the pulse envelope \cite{Bergmann1998}.
To examine the robustness of the controls against the error of the pulse profile, we multiply the pump field and the Stokes field by $\lambda_{\rm p}$ and $\lambda_{\rm S}$, respectively.
Figures~\ref{stability}(a),~\ref{stability}(b) and~\ref{stability}(c) show the dependence of the fidelity defined by the population of $|3,\downarrow\rangle$ at $t=T_2$ on $\lambda_{k}$ for the $\pi$-pulse control, the spin-STIRAP and the invariant-based engineering protocol, respectively.
It is seen that the spin-STIRAP and the invariant-based engineering protocols are more robust against the error of the pulse area compared to the $\pi$-pulse control.
The spin-STIRAP shows its robustness even when the pulse area is considerablly large as shown in Fig.~\ref{stability}(b).
The invariant-based engineering protocol is more robust than the $\pi$-pulse control especially for the case of $\lambda_{\rm p} \simeq \lambda_{\rm S}$ as shown in Fig.~\ref{stability}(c).
Figure~\ref{stability}(d) shows the dependence of the fidelity of the $\pi$-pulse control and the invariant-based engineering protocol for $\lambda_{\rm p} =\lambda_{\rm S}$.
The robustness of the invariant-based engineering protocol depends on the parameter $\delta$.
Note that the fidelity for $\lambda_{\rm p} =\lambda_{\rm S}=1$ is slightly less than unity
in Figs.~\ref{stability}(c) and \ref{stability}(d). 
This imperfection of the fidelity is attributed to the fact mentioned in the last paragraph of Sec.~\ref{Shortcuts to adiabaticity}.
Interestingly, the peak of fidelity appears at slightly away from $\lambda_{\rm p}=\lambda_{\rm S}=1$, and the value of $\lambda_{\rm p/S}$ corresponding to the peak depends on $\delta$ as seen in Fig.~\ref{stability}(d), although the detailed study of this dependence is beyond the scope of this paper.

\begin{figure}[h!]
\begin{center}
\includegraphics[width=7cm]{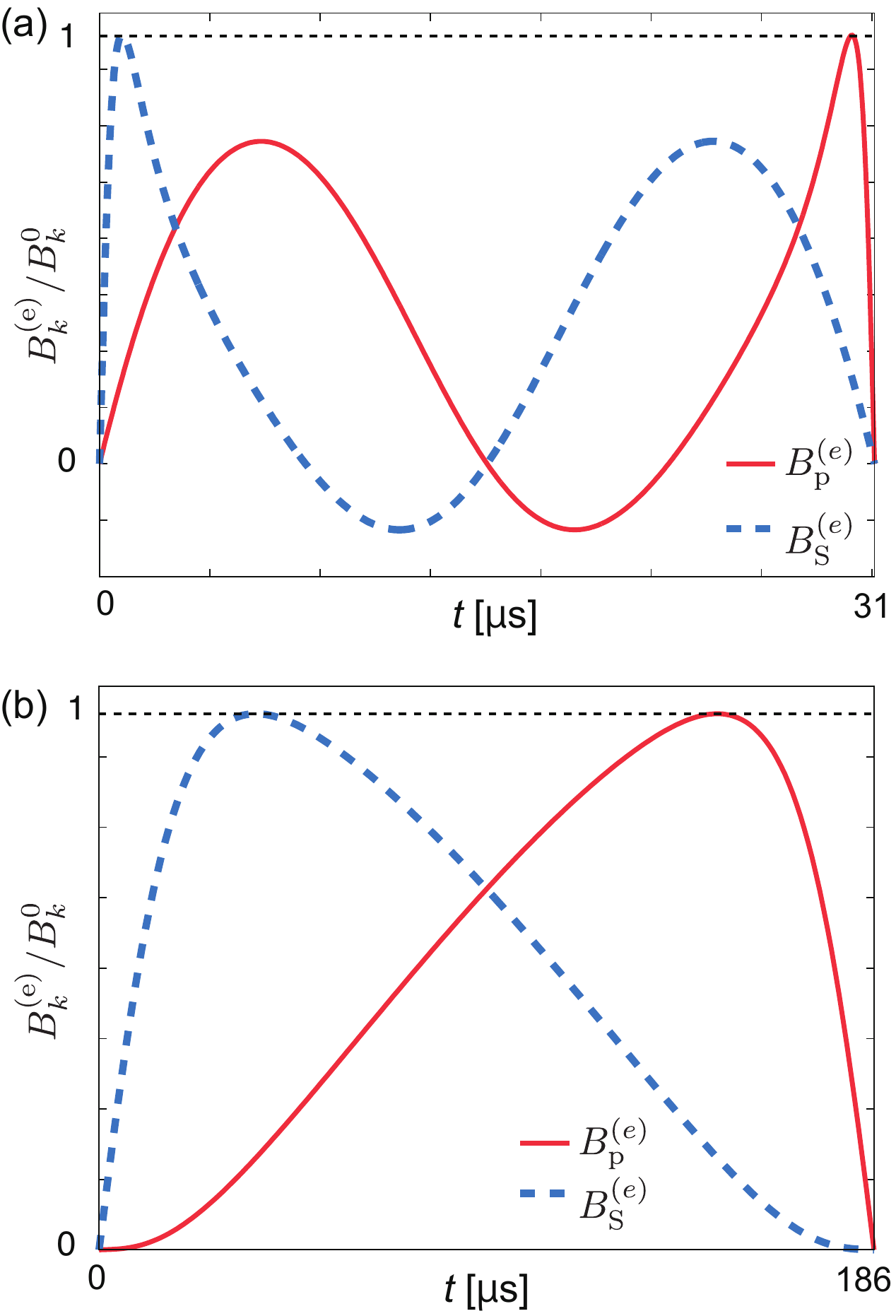}
\end{center}
\caption{(Color online) Envelop functions of the magnetic fields $B_{\rm p}^{\rm (e)}$ and $B_{\rm S}^{\rm (e)}$ normalized by $B_{\rm p}^{0}$ and $B_{\rm S}^{0}$, respectively, for
invariant-based engineering protocol for (a) $T_2-T_1=31$ $\mu$s, $\delta=0.5\pi$, $\varepsilon=0.02$  and (b) $T_2-T_1=186$ $\mu$s, $\delta=0.02\pi$ and $\varepsilon=0.02$. 
The other parameters used are shown in the caption of Table~\ref{table_STEP2}. }
\label{Bx_10_20_17}
\end{figure}

Now we show the results of the invariant-based engineering protocol with shorter and longer duration than the ones shown in Fig.~\ref{rabi_com}(c) and Fig.~\ref{pop}(c) to show that the invariant-based engineering protocol interpolates between the feature of other two controls.
We refer the invariant-based engineering protocol with the duration of 31~$\mu$s, $186$~$\mu$s and 62~$\mu$s, as short STA, long STA and medium STA for ease of expression.
The medium STA corresponds to $B_{\rm p,S}^{\rm (e)}$ in Fig.~\ref{rabi_com}(c).
We chose $\delta$ for short and long STAs such that the maximum value of the envelop functions is the same as Fig.~\ref{rabi_com}(c), while $\varepsilon$ is the same as Fig.~\ref{rabi_com}(c).
Figure \ref{Bx_10_20_17}(a) shows the envelop functions of the magnetic fields $B_{\rm p}^{\rm (e)}$ and $B_{\rm S}^{\rm (e)}$ for short STA.
The amplitude of the pump (Stokes) field increases in the first (second) half of step~II, respectively, compared to those in Fig.~\ref{rabi_com}(c).
Note that the order of these major peaks of $B_{\rm p}^{\rm (e)}$ and $B_{\rm S}^{\rm (e)}$ is the same as $\pi$-pulse control.
Figure \ref{Bx_10_20_17}(b) shows the envelop functions of the magnetic fields for long STA than Fig.~\ref{rabi_com}(c).
The profiles of the envelop functions are similar to those of the STIRAP control rather than the $\pi$-pulse control.

\begin{figure}[h!]
\begin{center}
\includegraphics[width=7cm]{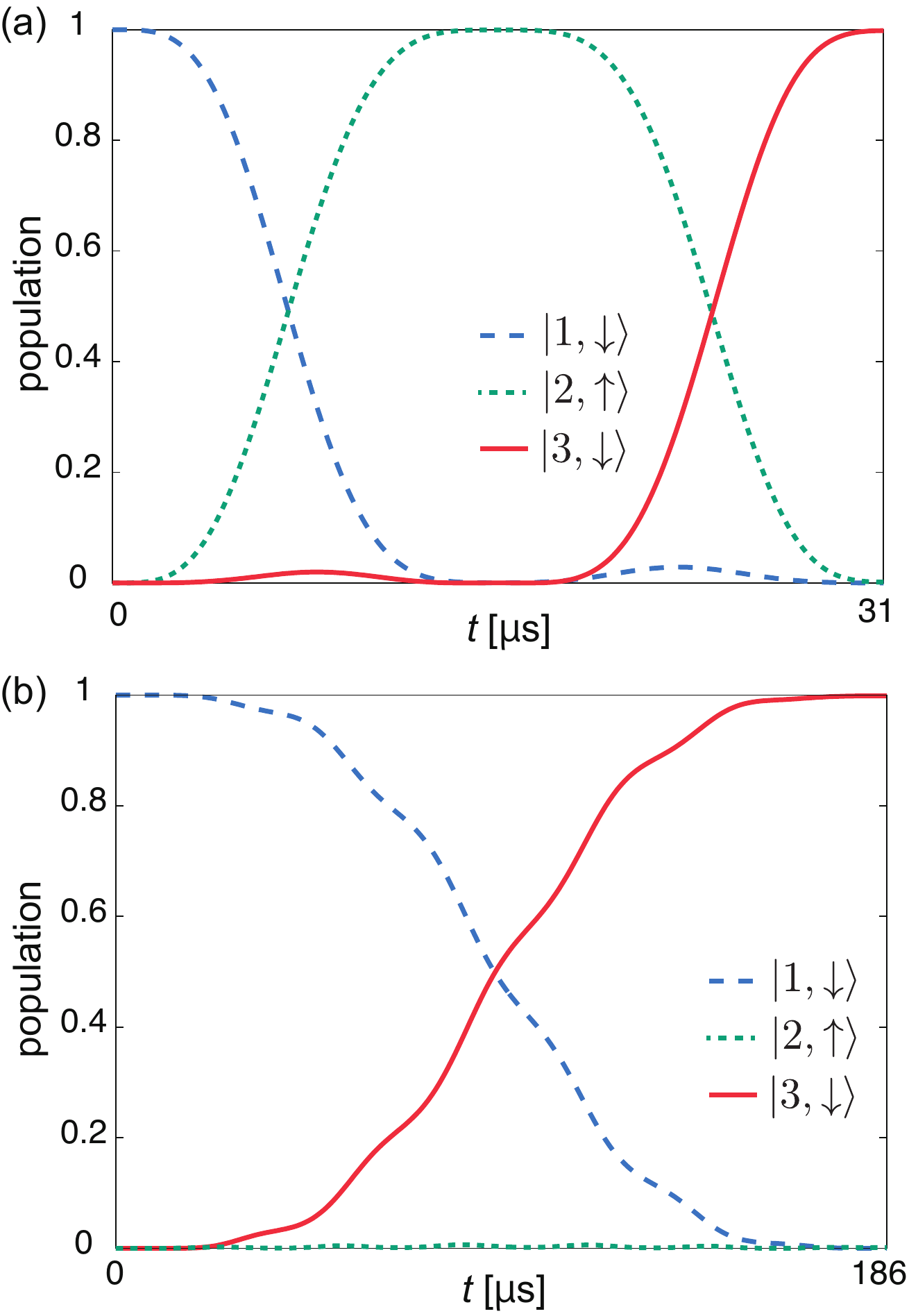}
\end{center}
\caption{(Color online) Time-dependence of the populations in the
invariant-based engineering protocol for (a) $T_2-T_1=31$ $\mu$s, $\delta=0.5\pi$ and $\varepsilon=0.02$  and (b) $T_2-T_1=186$ $\mu$s, $\delta=0.02\pi$ and $\varepsilon=0.02$.  
The other parameters used are shown in the caption of Table~\ref{table_STEP2}. }
\label{pop_10_20_17}
\end{figure}

Figures \ref{pop_10_20_17}(a) and \ref{pop_10_20_17}(b) show the time-dependence of populations corresponding to $B_{\rm p,S}^{\rm (e)}$ in Figs.~\ref{Bx_10_20_17}(a) and \ref{Bx_10_20_17}(b), respectively.
The feature of the populations in short STA is similar to that in the $\pi$-pulse control in Fig.\ref{pop}(a).
On the other hand, in long STA in Fig.~\ref{pop_10_20_17}(b), the population of state is directly transferred from $|1,\downarrow\rangle$ to $|3,\downarrow\rangle$ as it is in the STIRAP control in Fig.\ref{pop}(b).

\begin{figure}[h!]
\begin{center}
\includegraphics[width=7cm]{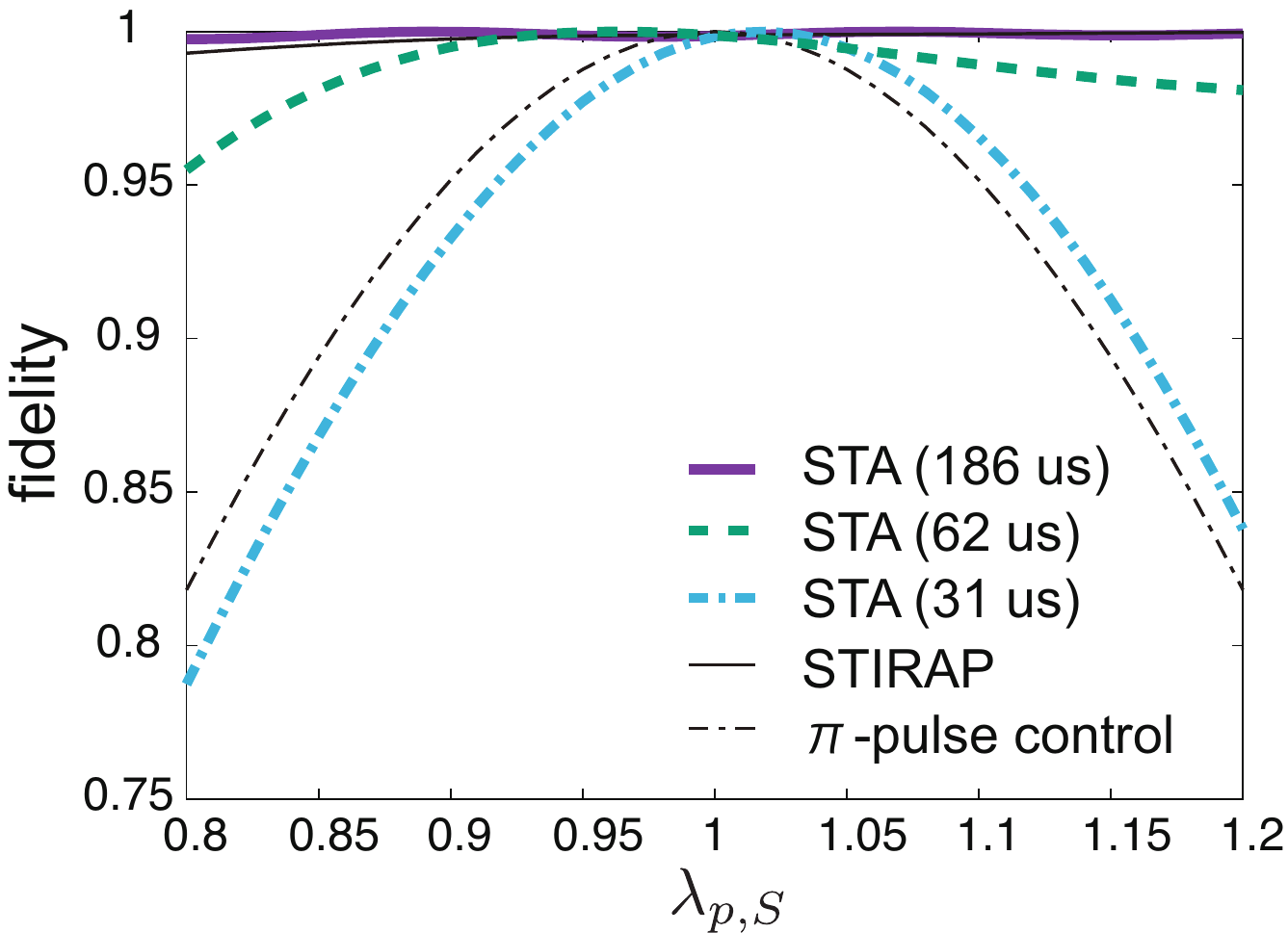}
\end{center}
\caption{(Color online) The $\lambda_{\rm p}$ dependence of the fidelity of the invariant-based engineering protocol for  the durations indicated in the panel, where $\lambda_{\rm p} =\lambda_{\rm S}$. 
The fidelities of the $\pi$-pulse control and the STIRAP control corresponding to Fig.~\ref{rabi_com}(a) and Fig.~\ref{rabi_com}(b), respectively, are also shown as functions of $\lambda_{\rm p}$ for comparison.}
\label{robustness_11_21_17}
\end{figure}

To show the trade-off between speed and robustness of the invariant-based engineering protocol,
we show the fidelity of short, long and medium STAs as function of $\lambda_{\rm p,S}$ in Fig.~\ref{robustness_11_21_17}.
The robustness of short STA is similar to the $\pi$-pulse control, while the robustness of long STA is considerably high as the STIRAP control.
It is even slightly higher than that of the STIRAP control because the effective duration of the STIRAP with the used parameters, when the population transfer actually occurs, is shorter than that of long STA.
The fidelity of medium STA is between short and long STA.
Thus, the invariant-based engineering protocol interpolates the feature of the $\pi$-pulse and the STIRAP controls.

The fluctuation of the potential of Dot~4 can degrade the control efficiency of step~II.
We consider the fluctuation of $V_4$ in Eq.~(\ref{eqV4}).
The change in the energy level intervals due to the change in $V_4$ is about one magnitude smaller than the change in $V_4$ (see Appendix~\ref{Influence of potential modulation to energy-level interval}).
To examine the influence of the potential fluctuation we introduce the fluctuation of the energy levels $\delta E_i$,
where $\delta E_i$ is the modulation of the energy from the value without potential fluctuation.
We model the fluctuation, $\delta E_i$, as a noise with a Gaussian distribution with the standard deviation $\sigma=k_{\rm B}T/10$ and time autocorrelation function $\alpha$.
The fluctuations of the energy levels are assumed to be independent of each other for simplicity.
The time evolution of the populations is calculated by solving the time-dependent Schr\"odinger equation with a fourth-order Runge-Kutta integrator with the time step of approximately 1~fs.
The other parameters used are shown in Table~\ref{table_STEP2}.
In the case with the level fluctuations, the total population of the states decreases to approximately 0.99 because of numerical error. The fidelity is defined by the renormalized population of $|3,\downarrow\rangle$.
It is seen that the fidelity of the three protocols is higher than 0.995 for $T=100$ mK and $\alpha = 1$ ps.

\subsection{Step IV}
\label{Step IV}
\begin{figure}
\begin{center}
\includegraphics[width=7.5cm]{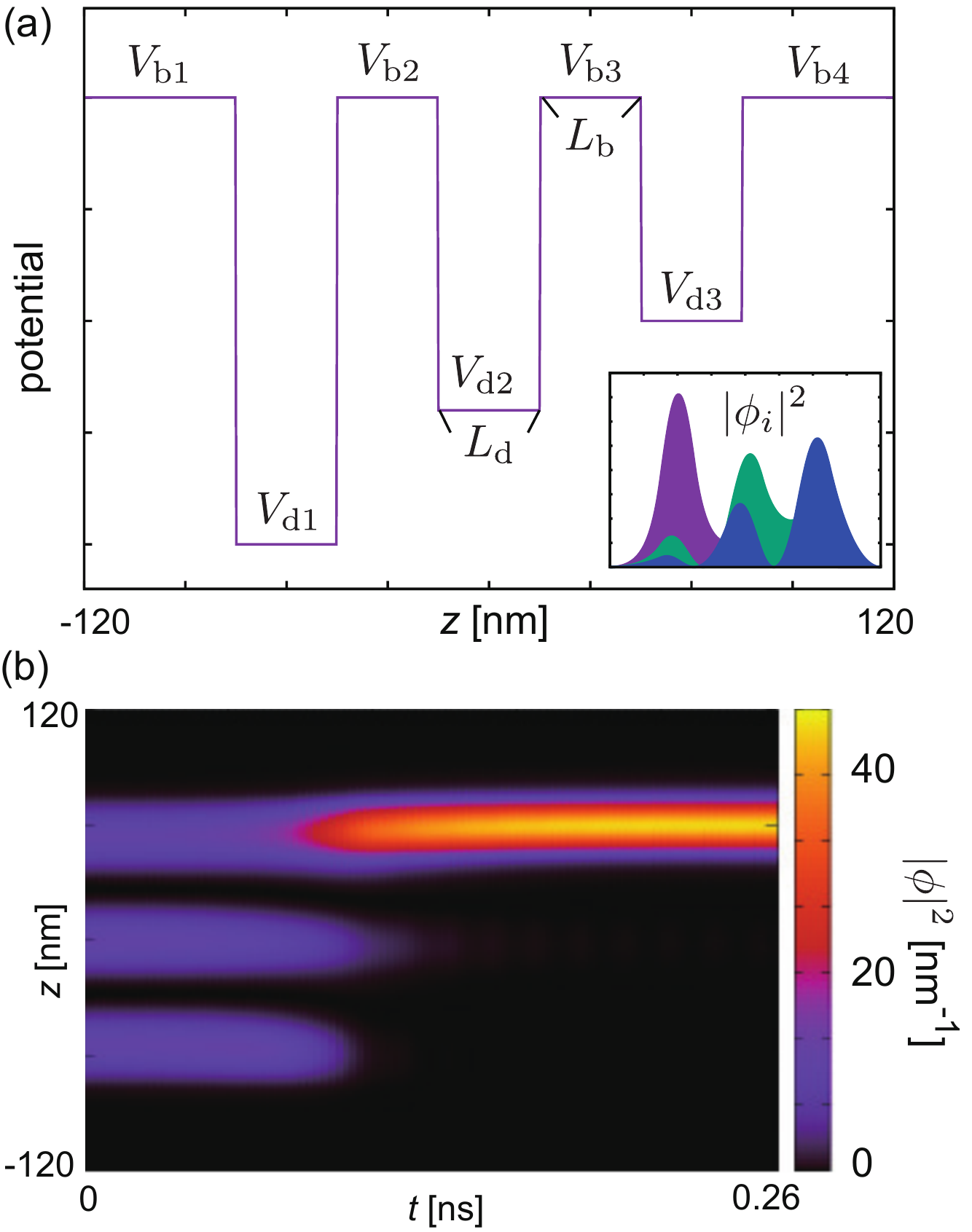}
\end{center}
\caption{(Color online) (a) Schematics of the one-dimensional model of the in-line dots. 
The height of the potential barriers and the depth of the potential wells are denoted by $V_{{\rm b}i}$ and $V_{{\rm d}j}$ for $i=1,2,3,4$ and $j=1,2,3$, respectively.
In this panel the difference in $V_{{\rm d}j}$ is exaggerated for ease of explanation.
The inset schematically shows the square of the amplitude of the wave function of the instantaneous eigenstates. The purple, green and blue colors correspond to $\phi_1$, $\phi_2$, and $\phi_3$, respectively.
(b) The time-evolution of the square of the amplitude of the wave function for the parameter set:
$L_{\rm d}=L_{\rm b}=30$~nm, $V_{\rm d1}^{(0)}=0$, $V_{\rm d2}^{(0)}=179$~$\mu$eV, $V_{\rm d3}^{(0)}=299$~$\mu$eV and 
$V_{\rm bi}^{(0)}=3.6$~meV. The initial state is $|3,\downarrow\rangle$.}
\label{potential_ph_4}
\end{figure}

To simulate step IV, we consider the one-dimensional model of the in-line dots with the rectangular potential illustrated in Fig.~\ref{potential_ph_4}(a).
Here, $V_{{\rm b}i}$ for $i=1,2,3,4$ are the barrier heights, and $V_{{\rm d}j}$ for $j=1,2,3$ are the potential depth of the dots.
The width of the dots $L_{\rm d}$ and the width of the barriers $L_{\rm b}$ are all taken to be $30$ nm.
Here, we use rectangular potentials for the dots for simplicity. 
Note that the details of the form of the potential do not influence the result because this step is based on adiabatic dynamics.
At the initial time of step IV, $t=T_3$, we take $V_{\rm b2,b3}=V_{{\rm d}j(=1,2,3)}=0<V_{\rm b1}=V_{\rm b4}$ so that the three dots are combined to form a single larger dot.
The middle barrier heights $V_{{\rm b}i} $ and dot potentials $V_{{\rm d}i}$ for $i=2,3$ are adiabatically increased 
from 0 to $V_{{\rm b}i}^{(0)}$ and $V_{{\rm d}i}^{(0)}$, respectively, as
\begin{eqnarray}
V_{{\rm b}i,{\rm d}i} = V_{{\rm b}i,{\rm d}i}^{(0)} [R(t-T_3)], \hspace{1cm} T_3\le t \le T_4
\end{eqnarray}
with $R(t) = [t-\sin(\omega_4 t)/\omega_4]/(T_4-T_3)$ and $\omega_4 = 2\pi/(T_4-T_3)$,
while the other parameters are kept constant, $V_{\rm b1,b4}=V_{\rm b1,b4}^{(0)}$ and $V_{\rm d1}=0$, where $T_4$ is the final time of step IV.
$R$ interpolates between 0 to 1 smoothly with $R(T_3)=0$, $R(T_4)=1$ and $R'(T_3)=R'(T_4)=0$, where a prime denotes time derivative.
Figure~\ref{potential_ph_4}(b) shows the time-evolution of the square of the amplitude of the wave function
for the initial state $|3,\downarrow\rangle$, which is the second excited state with spin down in the combined in-line dots.
It is seen that the wave function is mostly localized at Dot 3 at $t=T_4$.
The fidelity of step~IV defined by the overlap between the state at $t=T_4$ and the target energy eigenstate, is higher than 0.9999 for $T_4-T_3>0.26$~ns, thus unwanted non-adiabatic transitions are negligible.

Similarly to step~IV, we estimated the duration of steps~I and III. 
The fidelity of steps~I and III is higher than 0.9999 if the duration is longer than 2~ns (see Appendix~\ref{Step I and step III} for details).
Thus, the duration of steps~I, III, IV is much shorter than that for step~II.
%The wave function locates in dot 1 at $t=0$ and distributes also in dot 2 at $t=T_1$.
%We confirmed that the state remains in the same instantaneous eigenstate, and unwanted non-adiabatic transitions are negligible for $T_1=0.5$ ns.
%Non-adaibatic excitation is negligible.
%\begin{table}
%\begin{tabular}{cccccccc}
%\hline
%$L_{\rm b/d}$  & $V_{\rm d1}^{(0)}$ & $V_{\rm d2}^{(0)}$ & $V_{\rm d3}^{(0)}$ & $V_{\rm b1/2/3/4}^{(0)}$\\
%\hline
%30 nm & 0 & 179 $\mu$eV & 299 $\mu$eV  & 3.6 $m$eV   \\
%\hline   
%\end{tabular}
%\caption{Parameters of the system.}
%\label{table_pA}
%\end{table}

\section{Two-electron transport and nonlocal operation}
\begin{figure}
\begin{center}
\includegraphics[width=8cm]{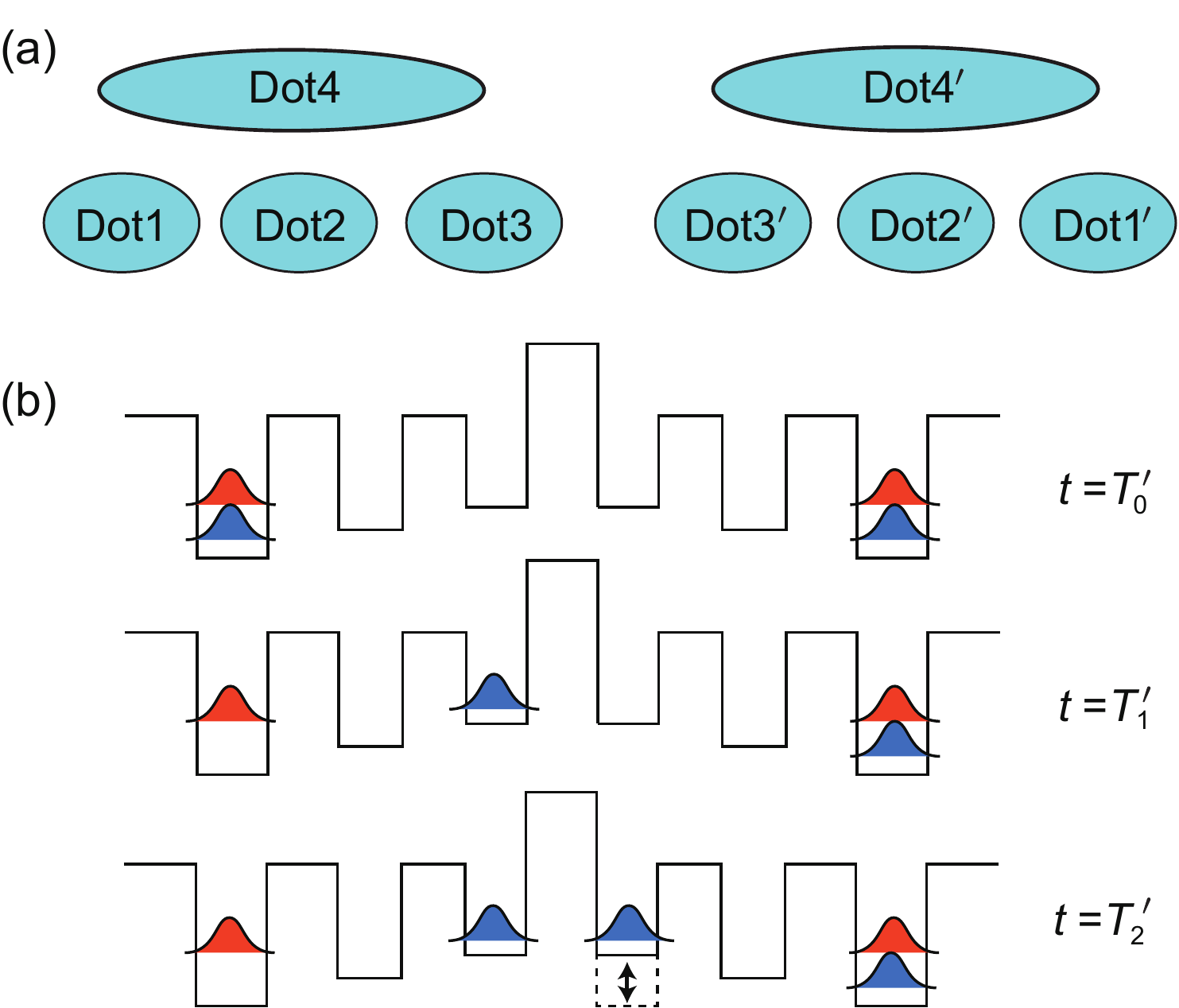}
\end{center}
\caption{(Color online) (a) Schematics of the system. The conducting lead for the AC current is not depicted here.
Dot 4 and Dot 4$'$ are used for transport of the left electron and the right electron, respectively.
(b) Schematics of the two-qubit gate operation. 
The black lines represent the potential of Dot~$i$ and Dot~$i'$ for $i=1,2,3$.
The blue and red color represent the square of the amplitude of the wave function of the spin-down and -up electrons, respectively
at the initial time $t=T_0'$, the end of the transfer of the left electron $t=T_1'$; and the end of the transfer of the right electron $t=T_2'$. 
The electrons take a superposition of spin-up and spin-down states.
The right electron is transferred to Dot~3$'$ only if the both electron spins are down initially.
The arrows at Dot~3$'$ represent the modulation of the dot potential by pulsing the gate voltage.}
\label{dot_lead_2electrons_7_28_16}
\end{figure}

We consider two-electron spin-selective transfer in the system depicted in Fig.~\ref{dot_lead_2electrons_7_28_16}(a)
to show that our protocol may be applicable to implement two-qubit gates.
We assume that an electron is trapped in Dot~1 and another electron is trapped in Dot~1$'$ at the initial time. 
The barrier potential between Dots~3 and 3$'$ is sufficiently high so that the two electrons cannot pass each other.
We aim to transfer the right electron to Dot~3$'$ only when both electron spins are initially down.

The schematics of the spin-selective electron transfer is shown in Fig.~\ref{dot_lead_2electrons_7_28_16}(b).
We first transfer the left electron to Dot~3.
We can use the method of the single electron transfer introduced above by taking into account the modulations of the overlapping factors and the resonance frequencies of the pump and the Stokes fields caused by the interaction between the electrons.
The left electron is transferred only if the electron spin is down.

Now we consider the right electron transfer.
First, the right electron is adiabatically transferred to Dot~4$'$.
When the left electron is trapped in Dot~3 or equivalently if the left electron spin is initially down, the effective potential for the right electron is deformed due to the interaction with the left electron in Dot~3 more than the case in which the left electron is trapped in Dot~1 or equivalently its spin is initially up.
%As illustrated in Fig.~\ref{potential_sin} for the single electron system, the wave function of higher levels distributes in wider range. 
%Due to the different distribution of the wave functions, each level is modulated differently by the interaction with the left electron. 
%This modulation is strong when the left electron is at Dot 3 compared to Dot 3.
Thus, the resonance frequencies of the pump and the Stokes fields depend on the initial spin of the left electron. 
The resonance frequencies for population transfer of the right electron in Dot~$4'$ depend on the occupancy of the left electron in Dot~3. The resonance frequencies, $f_{\rm p}$ and $f_{\rm S}$, for the right electron in Dot~$4'$ are modulated approximately by 900~MHz and 700~MHz, respectively, for $r_0=1.4~L_{\rm 4}$ and $\theta_0\simeq \pi/3 $ when Dot~3 is 317~nm distant from the center of Dot~4$'$. (See Appendix~\ref{Resonance frequency modulation due to Coulomb interaction} for this estimation.) The change of the overlapping factors is approximately 3 \% compared to the case when the Coulomb interaction is negligible. 
This property allows us to transfer the right electron depending on the initial set of the spins.
Again, we can use either of the non-adiabatic and the adiabatic schemes to transfer the right electron to Dot~3$'$.

This multi-electron transfer scheme can be used for non-local operations of qubits.
In the end of the control discussed above, the right electron is trapped in Dot~3$'$ only if both of the initial electron spins are down.
For example, pulsing the gate voltage of Dot~3$'$ can realize the non-local control of the phase of the qubits' state. The phase is tuned by the pulse intensity and the duration.
The electrons are brought back to the original dots (Dot~1 and Dot 1$'$) by the inverse electron transfer process.

The spin dependent phase in our scheme can be employed to implement important two-qubit gates, including the
CNOT (Controlled NOT) gate and the CZ (Controlled Phase) gate, for quantum circuit design \cite{Nielsen2000,Nakahara2008}. 
Let us denote $|\uparrow\rangle$ and $|\downarrow\rangle$ as $|0\rangle$ and $|1\rangle$, respectively.
The above protocol implements the gate 
\begin{eqnarray}
U_\varphi = \left( \begin{array}{cccc}
1& 0 & 0 & 0  \\
0 & 1 & 0 & 0  \\
0 & 0 & 1 & 0 \\
0 & 0 & 0 & e^{i\varphi}
\end{array}
 \right),
 \label{gate}
 \end{eqnarray}
up to an overall phase that we can safely ignore.
Here, $\varphi$ is the phase acquired when both electrons are in the spin sate $|1\rangle$, 
and the basis vectors are arranged in the order of $|00\rangle$, $|01\rangle$, $|10\rangle$ and $|11\rangle$. 
The phase is controlled by manipulating the gate voltage of Dot 3$'$.
Suppose that we adjust the parameters so that $\varphi=\pm\pi$.
Then, we obtain the CZ gate
\begin{eqnarray}
U_{\rm CZ} = |0\rangle\langle 0| \otimes I + |1\rangle\langle 1| \otimes \sigma_z.
\end{eqnarray}
The CNOT gate is obtained by applying the Hadamard gate before and after the operation of the CZ gate,
 \begin{eqnarray}
U_{\rm CNOT} &=& (I\otimes H_{\rm Had}) U_{\rm CZ} (I\otimes H_{\rm Had}) \nonumber\\
&=& |0\rangle\langle 0| \otimes I  + |1\rangle\langle 1| \otimes \sigma_x,
\end{eqnarray}
where
\begin{eqnarray}
H_{\rm Had} = \frac{1}{\sqrt{2}}\left( \begin{array}{cc}
1& 1   \\
1 & -1 
\end{array}
 \right)
 \label{Hgate}
 \end{eqnarray}
 is the Hadamard gate.
 We have demonstrated that our scheme implements the universal set of gates.
 
\section{Conclusion}
We have proposed the spin-selective, coherent electron transfer in quantum dot array.
The gradient of the oscillating magnetic field and the gate voltage control are utilized to separate the electron wave function into different quantum dots depending on the electron spin.
We have examined three different protocols: the non-adiabatic $\pi$-pulse control, the spin-STIRAP and the invariant-based engineering protocol.
The $\pi$-pulse control offers fast transport, and the spin-STIRAP offers a robust control against the error of the pulse area of the control field although the manipulation time is longer than the $\pi$-pulse control.
The invariant-based engineering protocol interpolates the other two protocols in the sense that it is faster than the spin-STIRAP and is more robust than the $\pi$-pulse control.
We also studied the robustness of the controls to the potential fluctuation.
This spin-selective electron transfer can be used for quantum non-demolition measurement of electron spin if it is followed by a measurement of its position which does not require absorption of the electron. 
This scheme can be extended to multi-electron systems offering the selectivity of the transport with respect to the set of spins, and can be used for non-local phase manipulation of the electrons including the CZ gate and the CNOT gate.

\section*{Acknowledgments}
We thank M. M\"{o}tt\"{o}nen, T. Otsuka, K. Takeda, J. Yoneda and K. Koshino for useful discussions.
M.N. is grateful to JSPS for partial support from a Grants-in-Aid for Scientific Research (Grant No. 26400422).
K. Y. T. acknowledges the support from the Academy of Finland (Grant No. 276528, 308161 and 314302).
S. M. acknowledges the support from JST ERATO (Grant No. JPMJER1601).

\appendix

\section{Derivation of $H_{\rm RWA}$}
\label{HRWA_ap}
Here we show the derivation of $H_{\rm RWA}$ in Eq.~(\ref{HRWA1}).
We consider an electron in 
Dot~4 governed by the Hamiltonian:
\begin{eqnarray}
H = \frac{{\bf p}^2}{2m^*} + V({\bm r}) +  \frac{g\mu_{\rm B}}{\hbar}{\bm B}(t,{\bm r})\cdot{\bm S},
\label{Hami_10_22_17}
\end{eqnarray}
where $m^*$ is the effective electron mass, ${\bm S}$ is the electron spin, $V$ is the potential of Dot~4 and
the magnetic field is given by
\begin{eqnarray}
{\bm B}(t,{\bm r}) = {\bm B}_z + {\bm B}_{\rm p}(t,{\bm r}) + {\bm B}_{\rm S}(t,{\bm r}).
\end{eqnarray}
We consider the subset of states $\{ |1,\downarrow\rangle, |2,\uparrow\rangle,  |3,\downarrow\rangle \}$ consisting of the energy eigenstates for ${\bm B}_{k}=0$, where $k={\rm p,S}$.
As mentioned in Sec.~\ref{Non-adiabatic and adiabatic spin-selective level transfers}, ${\bm B}_{k}$ is space dependent, and their frequency is tuned  so that ${\bm B}_{\rm p}$ couples $|1,\downarrow\rangle$ and $|2,\uparrow\rangle$, and ${\bm B}_{\rm S}$ couples $|2,\uparrow\rangle$ and $|3,\downarrow\rangle$.
The Hamiltonian of the reduced system spanned by the subset of states is represented as
\begin{eqnarray}
H' = \sum_{s_1,s_2}|s_1\rangle \langle s_1| H |s_2\rangle \langle s_2|,
\end{eqnarray}
where $s_1$ and $s_2$ run over $\{ (1,\downarrow), (2,\uparrow), (3,\downarrow)\}$.
Matrix elements of $H'$ are represented as 
\begin{eqnarray}
[H']_{s_1,s_2} = \langle s_1| H' |s_2\rangle.
\label{Hprime_10_26_17}
\end{eqnarray}

Now we use a rotating frame by transforming the state as
\begin{eqnarray}
|\Psi_R\rangle = U_R
 |\Psi\rangle,
\end{eqnarray}
with $U_R$ defined by
\begin{eqnarray}
U_R = \left( \begin{array}{ccc}
e^{iE_{1,\downarrow}t/\hbar} & 0 & 0  \\
0 & e^{iE_{2,\uparrow}t/\hbar} & 0 \\
0 & 0 & e^{iE_{3,\downarrow}t/\hbar}
\end{array}
 \right),
\end{eqnarray}
where $|\Psi\rangle$ is the state evolving under $H'$.
The Schr$\ddot{\mbox{o}}$dinger equation of $|\Psi_R\rangle$ is represented as
\begin{eqnarray}
i\hbar\frac{\partial }{\partial t}|\Psi_R\rangle &=&  H_R|\Psi_R\rangle,
\end{eqnarray}
with
\begin{eqnarray}
H_R=
 i\hbar(\partial_t U_R)U_R^\dagger + U_RH'U_R^\dagger,
\label{SE_10_22_17}
\end{eqnarray}
where $\partial_t$ denotes time derivative.
Note that the diagonal elements of $H_R$ cancel out due to $i\hbar(\partial_t U_R)U_R^\dagger$, and also that
\begin{eqnarray}
\langle s_1 |  \Big{[}\frac{{\bf p}^2}{2m^*} + V({\bm r}) +  \frac{g\mu_{\rm B}}{\hbar}{\bm B}_z\cdot{\bm S}\Big{]} |s_2\rangle=0
\label{eq1_10_26_17}
\end{eqnarray}
for $s_1\ne s_2$ because $|s_1\rangle$ and $|s_2\rangle$ are energy eigenstates for ${\bm B}_{k}=0$.
Thus, only ${\bm B}_k$ can contribute to $H_R$.
Using Eqs.~(\ref{Bx1}), (\ref{Hprime_10_26_17}), (\ref{SE_10_22_17}) and (\ref{eq1_10_26_17}) we can obtain an off-diagonal element of $H_R$ as
\begin{eqnarray}
[H_R]_{1\downarrow,2\uparrow} &=& [H']_{1\downarrow,2\uparrow} e^{i(E_{1,\downarrow}-E_{2,\uparrow})t/\hbar},\nonumber\\
&=&\sum_k \langle 1,\downarrow| \frac{g\mu_{\rm B}}{\hbar}{\bm B}_k\cdot{\bm S} |2,\uparrow\rangle e^{i(E_{1,\downarrow}-E_{2,\uparrow})t/\hbar}\nonumber\\
&=&\sum_k\frac{g \mu_{\rm B}B_k^{\rm (e)}(t)}{2} \cos(\omega_k t) e^{i(E_{1,\downarrow}-E_{2,\uparrow})t/\hbar}\nonumber\\
&&\times  \int d{\bm r}\langle 1,\downarrow |{\bm r},\downarrow\rangle \eta({\bm r}) \langle {\bm r},\uparrow |2,\uparrow\rangle,
\end{eqnarray}
where the integral in the last line is $\mu_{\rm p}$ defined in Eq.~(\ref{mu_10_26_17}).
Noting that $\omega_{\rm p}=(E_{1,\downarrow}-E_{2,\uparrow})/\hbar$,
we can approximate $[H_R]_{1\downarrow,2\uparrow}$ with the RWA to obtain a matrix element of $H_{\rm RWA}$ as
\begin{eqnarray}
[H_R]_{1\downarrow,2,\uparrow} \simeq 
\frac{B_{\rm p}^{\rm (e)}(t) g \mu_{\rm B}\mu_{\rm p}}{4} =: [H_{\rm RWA}]_{1\downarrow,2,\uparrow}.
\end{eqnarray}

In the RWA we neglected terms rapidly oscillating compared to slowly changing elements due to the time dependent envelope function $B_{k}^{\rm (e)}$ assuming that the state is almost unchanged during a single period of such fast oscillation of Hamiltonian matrix elements, and the influence of the oscillations is canceled out.
The other matrix elements of $H_{\rm RWA}$ are obtained in the same manner.
Finally we obtain the effective Hamiltonian in Eq.~(\ref{HRWA1}).

The RWA is applicable when the AC magnetic fields are sufficiently small,  that is, the Rabi frequencies are much smaller than  $\delta E/\hbar$ as in our case,  where $\delta E$ is the energy interval between relevant levels (a few lowest levels in our case). On the other hand if external oscillating fields are too strong the above assumption is no longer valid. Degradation of the control efficiency due to too strong fields was studied, e.g , in Ref.~[\!\!\citenum{Jakubetz2012}].

\section{Another method of wave function splitting}
\label{Another method of wave function splitting}
We briefly discuss a simpler method of wave function splitting using a single $\pi$-pulse which couples $|1,\downarrow\rangle$ and $|2,\uparrow\rangle$, in contrast with our protocol which uses two $\pi$-pulses in order to compare with the STIRAP and the shortcut to adiabaticity protocols. Steps I, III and IV are the same as those in the main text. In step II, $|1,\downarrow\rangle$ is transffered to $|2,\uparrow\rangle$ by a single pulse. In step IV, $|2,\uparrow\rangle$ states is adiabatically carried to Dot~2, while $|1,\uparrow\rangle$ is trapped in Dot~1 in the end of the control. Thus, the wave function is split into different dots depending on the initial electron spin. Dot~3 is not used in this scheme.
This method can be also utilized for the two-qubit gates in the same manner as discussed in the main text.
Although this scheme is simple, it does not enjoy robustness against error, which spin-STIRAP and STA protocols offer.

%If we want to flip the spin of $|2,\uparrow\rangle$, we can use oscillating magnetic field after shuttling $|2,\uparrow\rangle$ wave function to other quantum dots so that $|1,\uparrow\rangle$ state does not feel the oscillating magnetic field.

\section{Influence of potential modulation to energy-level interval}
\label{Influence of potential modulation to energy-level interval}
\begin{figure}[h!]
\begin{center}
\includegraphics[width=7.5cm]{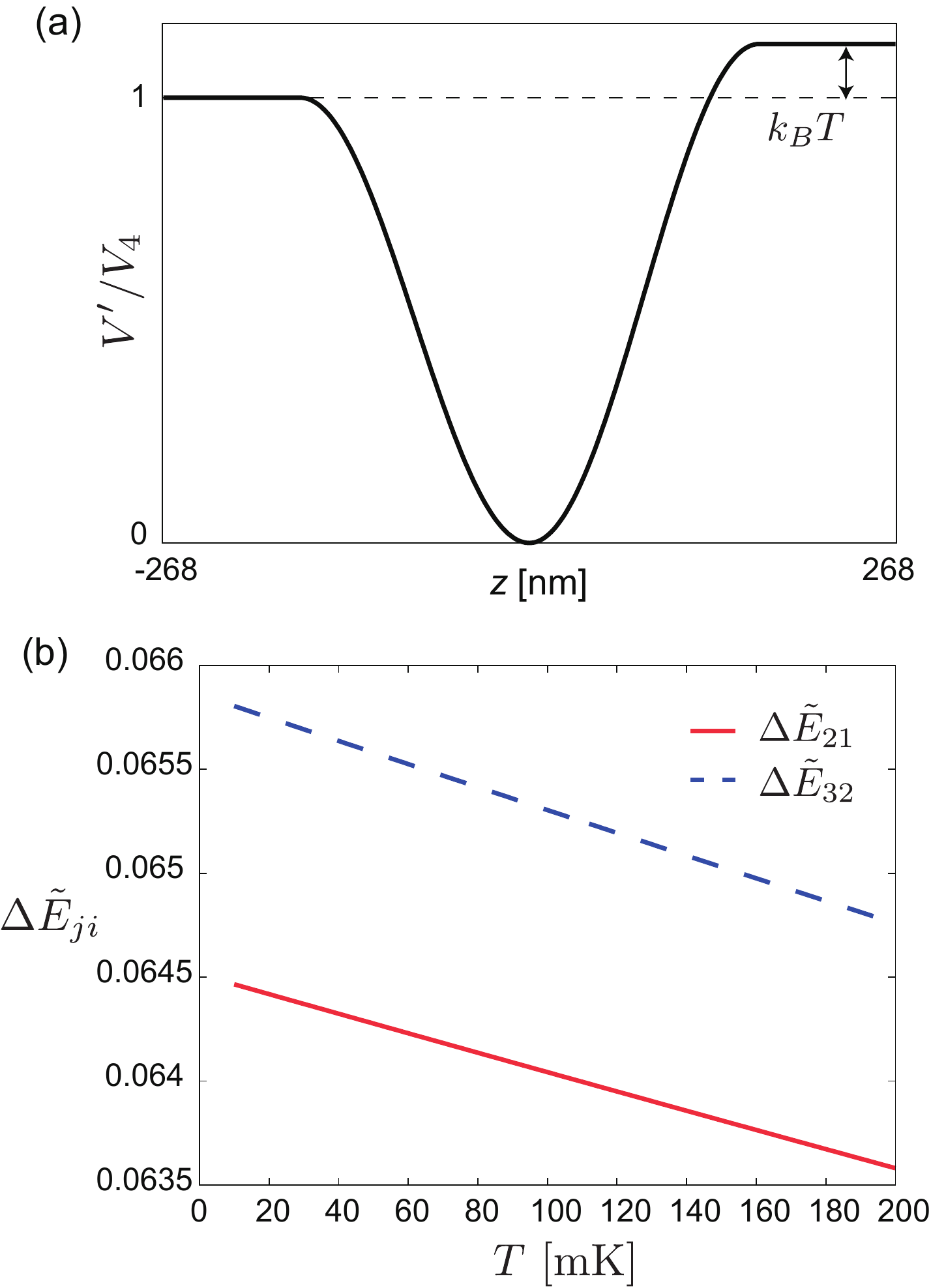}
\end{center}
\caption{(Color online) Schematics of the profile of $V'/V_4$ defined in Eq.~(\ref{eqV4_2}).
The temperature dependence of the change of energy level intervals normalized by $k_{\rm B} T$ for a few of lowest levels.}
\label{dif_de3}
\end{figure}

We consider the modulation of $V(z)$ from the original form in Eq.~(\ref{eqV4}).
The modulated potential $V'$ is represented as
\begin{eqnarray}
V'(z) = 
\begin{cases}
V(z) &  {\rm for} \ z < 0.\\
\Big{(}1+\frac{k_{\rm B} T}{V_4}\Big{)}V(z)  &  {\rm for} \ z \ge 0.
\end{cases}
\label{eqV4_2}
\end{eqnarray}
The potential wall of $V'$ is $k_{\rm B} T$ higher for positive $z$ than the other side as schematically depicted in Fig.~\ref{dif_de3}(a).
The interval between energy levels change from original one due to the potential modulation.
Now we consider the change of an energy interval from the original one defined by
\begin{eqnarray}
\Delta \tilde E_{ji} = \frac{(E_j'-E_i') - (E_j-E_i)}{k_{\rm B}T},
 \end{eqnarray}
where $E_i'$ and $E_i$ are eigenenergies corresponding to $V'$ and $V$, respectively.
The change of an energy interval is normalized by $k_{\rm B}T$.
Figure~\ref{dif_de3}(b) shows $\Delta \tilde E_{ji}$ for a few of the lowest levels.
It is seen that $\Delta \tilde E_{ji} < 0.1$ for 10~mK$<T<$200~mK.
The change of a level interval due to the change in $V_4$ is about one order of magnitude smaller than that of $V_4$.

\section{Resonance frequency modulation due to Coulomb interaction}
\label{Resonance frequency modulation due to Coulomb interaction}
\begin{figure}[h!]
\begin{center}
\includegraphics[width=7.5cm]{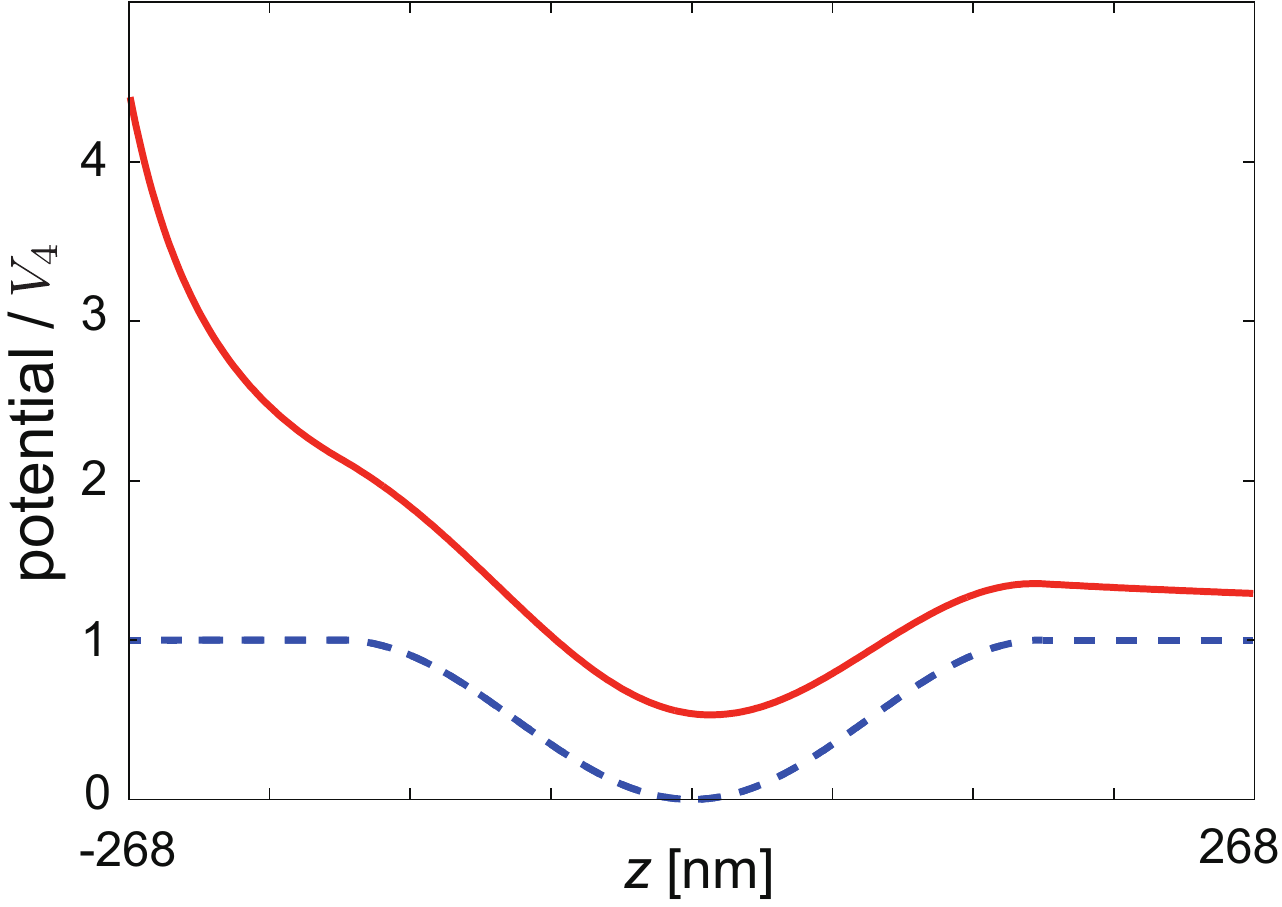}
\end{center}
\caption{(Color online) The effective potentials normalized by $V_4$ for the electron in Dot~$4'$. The solid curve and the dashed curve are for the case in which the left electron is located at $z=-l_z$ and for the case in which Coulomb interaction is negligible, respectively.}
\label{potential_com_5_16_17}
\end{figure}

The resonance frequencies for population transfer of the right electron in Dot~$4'$ depend on the position of the left electron (see Fig.~\ref{dot_lead_2electrons_7_28_16}). We consider the modulation of the resonance frequencies due to the Coulomb interaction between the electrons. 
We estimate the frequency modulation with a simple one-dimensional model although the actual modulation depends on the detail of device design such as the geometry  of electromagnetic components constituting of the quantum dots.
We assume that the center of Dot~$3$ is separated from the center of Dot~$4'$ by $l_z$.
The influence of the Coulomb interaction is approximated by the modulation of the potential of Dot~4$'$. Note that the exchange interaction can be neglected because there is no exchange of electrons across the center barrier which is set to be sufficiently high.
When the left electron is located at Dot~3 the right electron is subjected by the additional potential:
\begin{eqnarray}
\Delta V(z) = \frac{1}{4\pi \varepsilon_{\rm eff}}\frac{e^2}{|z-l_z|},
 \label{DeltaV_5_16_17}
 \end{eqnarray}
where $\varepsilon_{\rm eff}=11.68\varepsilon_{\rm 0}$ and we take the center of Dot~$4'$ as the origin.
We ignore the spatial distribution of the wave function of the left electron for simplicity.
On the other hand, the additional potential can be put to zero when the left electron is located in Dot~1 because it is sufficiently far or it can be carried to other dots which are sufficiently far from Dot~$4'$ if necessary. 
The effective potential $V(z)+\Delta V(z)$ is shown in Fig.~\ref{potential_com_5_16_17}.
When the left electron is in Dot~3 and $l_z=317$~nm, the resonance frequencies, $f_{\rm p}$ and $f_{\rm S}$ for Dot~4$'$, are approximately 900~MHz and 700~MHz higher, respectively, compared to the case when the Coulomb interaction is negligible. 
The parameters for the dot and $r_0$ and $\theta_0$ are the same as those used in Fig.~\ref{pop}.
When the left electron is in Dot~3, the overlapping factors, $\mu_{\rm p}$ and $\mu_{\rm S}$ are approximately 3~\% larger and 3~\% smaller, respectively, compared to the case when the Coulomb interaction is negligible. 

\section{Step I and step III}
\label{Step I and step III}
In step~I, we adiabatically transfer an electron to Dot~4 from Dot~1.
Here, we show that the duration of step~I can be much shorter than that of step~II.
For ease of estimation of the duration, we first make a combined dot composed of Dots~1--3 by lowering the barriers between them. 
(The required duration of this process is less than 1~ns as shown in Sec~\ref{Step IV} for the opposite process.)

\begin{figure}
\begin{center}
\includegraphics[width=7cm]{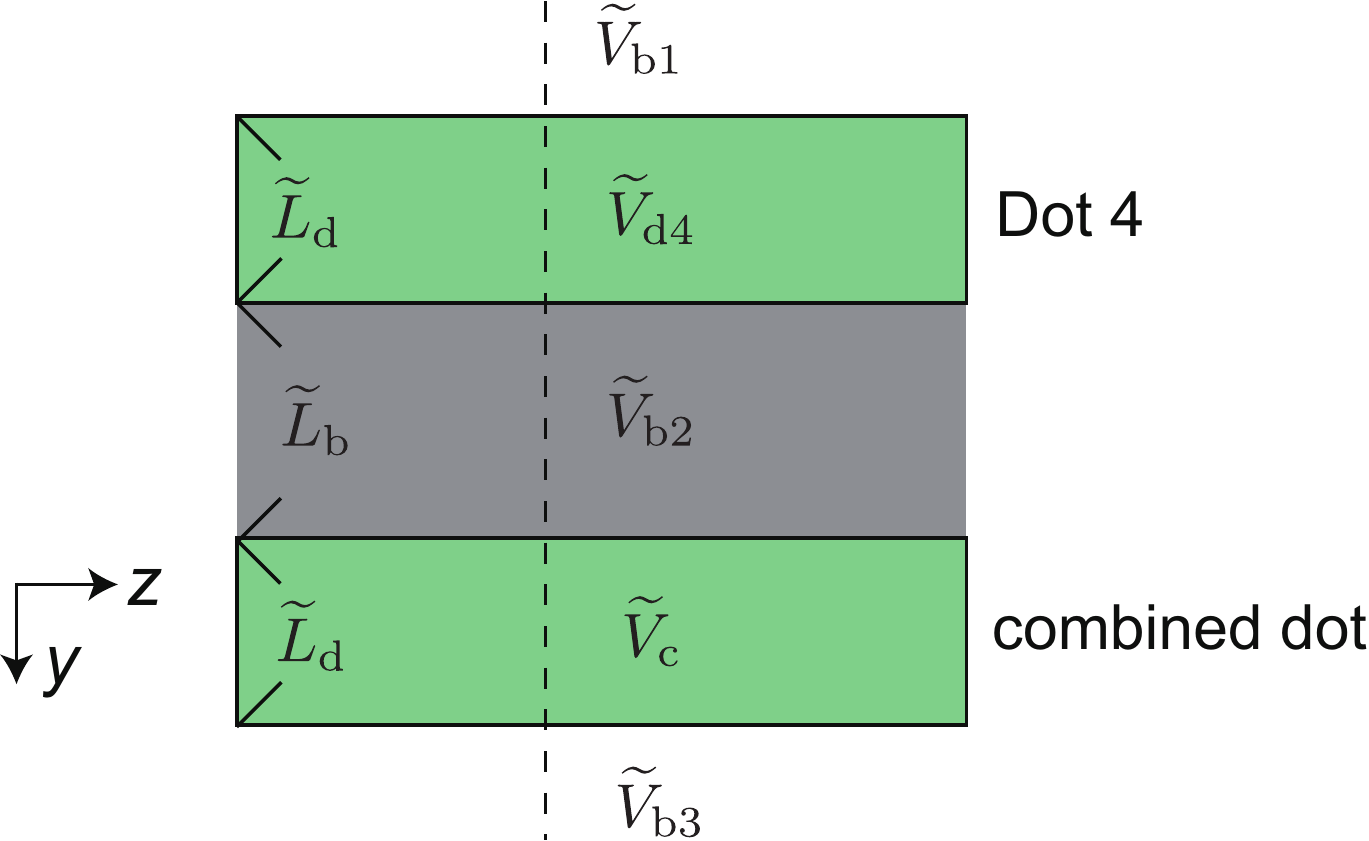}
\end{center}
\caption{(Color online) Schematics of Dot~4 (upper green square) and combined dot (lower green square) in step~I.
The gray square represents potential barrier between the dots.
$\widetilde{V}_{\rm bi}$ for $i=1,2,3$ is barrier potential, and $\widetilde{V}_{\rm d4}$ and $\widetilde{V}_{\rm c}$ are the potential of Dot~4 and the combined dot, respectively.
$\widetilde{L}_{\rm b}$ and $\widetilde{L}_{\rm d}$ are the width of the center barrier and the dots in the $y$-direction.
The potential profile along the dashed line is shown in Fig.~\ref{wf1_3_11_23_17}(a).
}
\label{stepI_10_24_17}
\end{figure}

The electron is trapped in the combined dot now.
Then we adiabatically transfer the electron to Dot~4.
We assume, for ease of analysis, that the combined dot and Dot~4 are rectangular in the $yz$-plane and of the same size as depicted in Fig.~\ref{stepI_10_24_17}.
The barrier potential outside the colored region in Fig.~\ref{stepI_10_24_17} is so high that the wave function is almost vanishing there.
For the transfer of the electron we gradually change the potential  $\widetilde{V}_{\rm c}$ of the combined dot, the potential  $\widetilde{V}_{\rm d4}$ of Dot~4 and the barrier height $\widetilde{V}_{\rm b2}$ between them.
The time dependence of the potentials are assumed as
\begin{eqnarray}
\widetilde{V}_{\rm c}(t) &=& \frac{\tilde{V}}{2}\Big{[}1 - \cos\Big{(}\frac{\pi t}{T_1}\Big{)}\Big{]},\nonumber\\
\widetilde{V}_{\rm d4}(t) &=& \frac{\tilde{V}}{2}\Big{[}1 + \cos\Big{(}\frac{\pi t}{T_1}\Big{)}\Big{]},\nonumber\\
\widetilde{V}_{\rm b2}(t) &=& \frac{\tilde{V}}{2}\Big{[}1 + \cos\Big{(}\frac{2\pi t}{T_1}\Big{)}\Big{]},
\end{eqnarray}
with constant $\tilde{V}$ for $0\le t\le T_1$, where we redefined $t=0$ as the time when the combined dot is formed for ease of notation.

\begin{figure}
\begin{center}
\includegraphics[width=7cm]{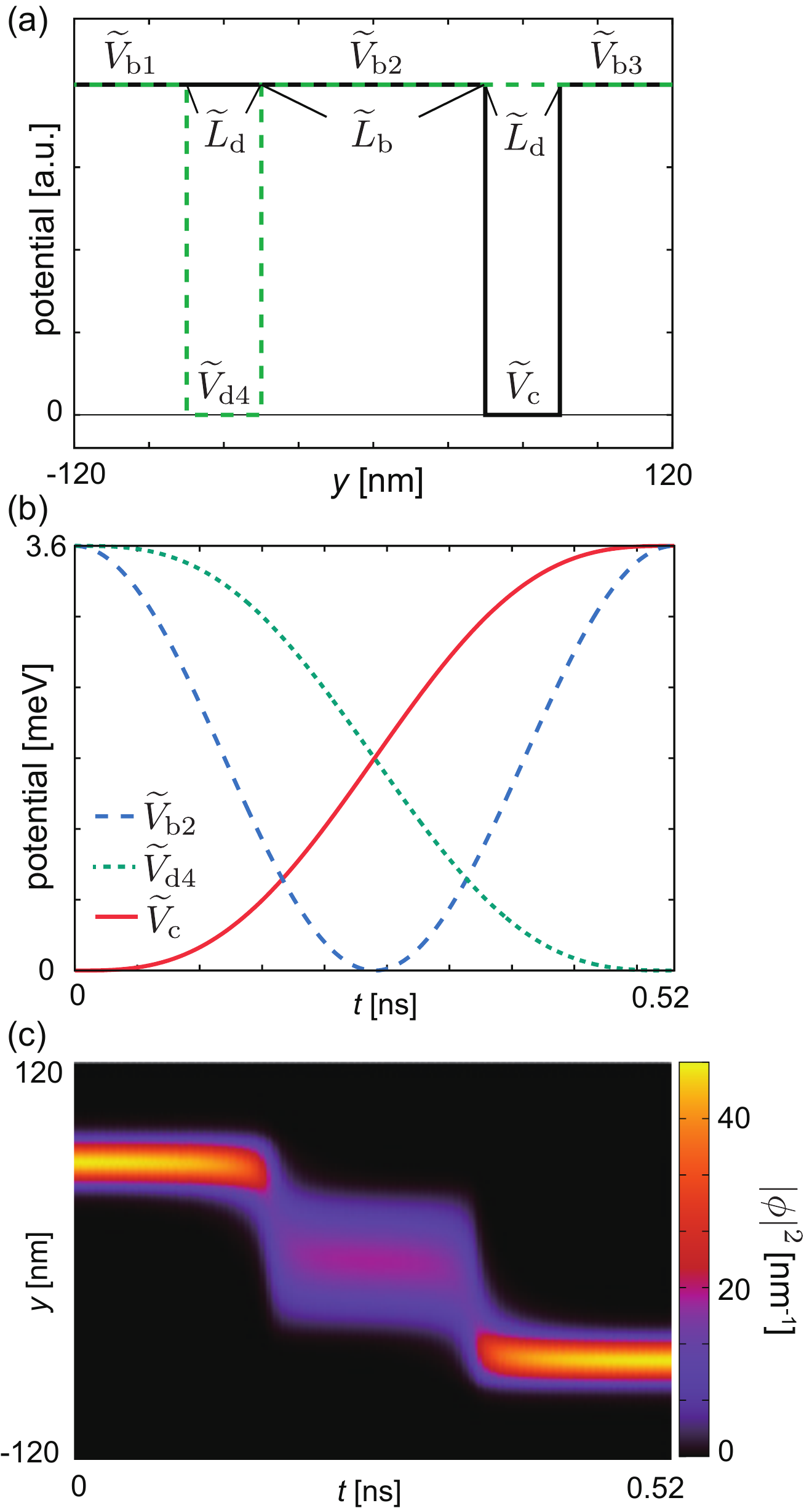}
\end{center}
\caption{(Color online) (a) Schematics of potential profile of the one-dimensional model for Dot~4 and the combined dot, where $\widetilde{L}_{\rm b}$ and $\widetilde{L}_{\rm d}$ are the width of the center barrier and the dots in the $y$-direction.
The thick solid and dashed lines are for the initial and final potential profiles of step~I, respectively, while the thin solid line indicates the zero of potential.
(b) Time dependence of $\widetilde{V}_{\rm b2}$, $\widetilde{V}_{\rm c}$ and $\widetilde{V}_{\rm d4}$
for $0<t<T_1$ for the parameter set: $\widetilde{V}_{\rm b1}=\widetilde{V}_{\rm b3}=\widetilde{V}=3.6$~meV,  $T_1=0.52$~ns, $\widetilde{L}_{\rm b}=90$~nm and $\widetilde{L}_{\rm d}=30$~nm.
(c) The time-evolution of the square of the amplitude of the wave function for the same parameter set as (b). }
\label{wf1_3_11_23_17}
\end{figure}

To simulate the dynamics of the electron for $0\le t\le T_1$ we use the one-dimensional model depicted in Fig.~\ref{wf1_3_11_23_17}(a) because
this system is separable with respect to $y$ and $z$ directions, and the $z$-dependence of the system is unchanged.
The potential profile in Fig.~\ref{wf1_3_11_23_17}(a) corresponds to the potential on the dashed line in Fig.~\ref{stepI_10_24_17}.
Here, $\widetilde{V}_{\rm b1}=\widetilde{V}_{\rm b3}$  are constant barrier height.
Figure~\ref{wf1_3_11_23_17}(b) shows the time dependence of  $\widetilde{V}_{\rm b2}$, $\widetilde{V}_{\rm c}$ and $\widetilde{V}_{\rm d4}$.
Figure~\ref{wf1_3_11_23_17}(c) shows the time-evolution of the square of the amplitude of the wave function which distributes in the combined dot initially.
The initial state is the ground state of the system.
The wave function is gradually moved to the center region as $\widetilde{V}_{\rm b2}$ is lowered and $\widetilde{V}_{\rm c}$ is raised, then it is gradually moved to Dot~4 as $\widetilde{V}_{\rm b2}$ is raised and $\widetilde{V}_{\rm d4}$ is lowered.
The fidelity of step~I, which is defined by the overlap between the state at $t=T_1$ (final time of step~I) and the target energy eigenstate, is higher than 0.9999 for $T_1>0.52$~ns for parameters shown in the caption of Fig.~\ref{wf1_3_11_23_17}.

Because step~III is the opposite control to step~I, the required duration is the same as step~I.
The required duration for steps~I and III are more than three orders of magnitudes shorter than step~II although the required duration depends on the detail of the profile and the time dependence of the potential more or less.
Therefore, step~II dominates the execution time of the total process.

\end{document}